%% file: ring_arxiv.tex
\label{lastpage}

\pdfoutput=1
\documentclass[onecolumn,a4paper,fleqn,usenatbib]{mnras}

\usepackage[T1]{fontenc}
\usepackage{ae,aecompl,soul}

\usepackage{graphicx}	
\usepackage{amsmath}	
\usepackage{amssymb}	

\usepackage{deluxetable}
\def\msol{\hbox{$\hbox{M}_\odot$}}
\def\lsol{\hbox{$\hbox{L}_\odot$}}

\renewcommand {\deg}   {\mbox{$^\circ$}}

\newcommand   {\kms}   {\mbox{km\,s$^{-1}$}}
\renewcommand {\ga}    {\mbox{\rlap{\hbox{\lower5pt\hbox{$\sim$}}}\hbox{$>$}}}
\renewcommand {\la}    {\mbox{\rlap{\hbox{\lower5pt\hbox{$\sim$}}}\hbox{$<$}}}

\title[High Velocity]{The interaction of a large-scale nuclear wind\\
                      with the high velocity HII region G0.17+0.15}

\author[Yusef-Zadeh, Zhao, Arendt, et al.]
{F. Yusef-Zadeh$^1$\thanks{E-mail: zadeh@northwestern.edu}, Jun-Hui Zhao$^2$,  R. Arendt$^3$, M. Wardle$^{4}$, M. Royster$^5$, L. 
Rudnick$^6$, \& J. Michail$^1$\\
$^{1}$Department of Physics and Astronomy Northwestern University, Evanston, IL 60208\\
$^{2}$Center for Astrophysics | Harvard-Smithsonian, 60 Garden Street, Cambridge, MA 02138, USA; jzhao@cfa.harvard.edu\\
$^{3}$UMBC/GSFC/CRESST 2, Code 665, NASA/GSFC, 8800 Greenbelt Rd, Greenbelt MD 20771\\
$^{4}$School of Mathematical and Physical Sciences,  Centre for Astronomy and \\
Space Technology, Macquarie University, Sydney NSW 2109, Australia\\
$^{5}$Department of Physics,  Sequaya Community College,  Visalia, CA 93277\\
$^{6}$Minnesota Institute for Astrophysics, University of Minnesota, 116 Church St. SE, Minneapolis, MN 55455, USA
}

\date{Accepted XXX. Received YYY; in original form ZZZ}

\pubyear{2019}
\begin{document}
\label{firstpage}
\pagerange{\pageref{firstpage}--\pageref{lastpage}}
\maketitle

\def\kms{km s$^{-1}$}
\def\Blos{B$_{\rm los}$}
\def\etal   {{\it et al.}}                     
\def\psec           {$.\negthinspace^{s}$}
\def\pasec          {$.\negthinspace^{\prime\prime}$}
\def\pdeg           {$.\kern-.25em ^{^\circ}$}
\def\degree{\ifmmode{^\circ} \else{$^\circ$}\fi}
\def\ut #1 #2 { \, \textrm{#1}^{#2}} 
\def\u #1 { \, \textrm{#1}}          
\def\nH {n_\mathrm{H}}
\def\ddeg   {\hbox{$.\!\!^\circ$}}              
\def\deg    {$^{\circ}$}                        
\def\le     {$\leq$}                            
\def\sec    {$^{\rm s}$}                        
\def\i      {\hbox{\it I}}                      
\def\v      {\hbox{\it V}}                      
\def\dasec  {\hbox{$.\!\!^{\prime\prime}$}}     
\def\asec   {$^{\prime\prime}$}                 
\def\dasec  {\hbox{$.\!\!^{\prime\prime}$}}     
\def\dsec   {\hbox{$.\!\!^{\rm s}$}}            
\def\min    {$^{\rm m}$}                        
\def\hour   {$^{\rm h}$}                        
\def\amin   {$^{\prime}$}                       
\def\lsol{\, \hbox{$\hbox{L}_\odot$}}
\def\sec    {$^{\rm s}$}                        
\def\etal   {{\it et al.}}                     
\def\la{\lower.4ex\hbox{$\;\buildrel <\over{\scriptstyle\sim}\;$}}
\def\ga{\lower.4ex\hbox{$\;\buildrel >\over{\scriptstyle\sim}\;$}}

\begin{abstract} 
We investigate the nature of a Galactic center source, G0.17+0.15, lying along the northern extension of the Radio Arc near 
$l\sim0.2^\circ$. G0.17+0.15 is an HII region located toward the eastern edge of the radio bubble, embedded within the highly polarized Galactic 
center eastern Lobe where a number of radio filaments appear to cross through the HII region. We report the detection of hydrogen and helium recombination 
lines with a radial velocity exceeding 140 \kms\, based on GBT and VLA observations.  The morphology of G0.17+0.15, aided by kinematics, and spectral 
index characteristics, suggests the presence  of an external pressure dragging and shredding the ionized gas.  We argue that this ionized cloud 
is interacting with a bundle of radio filaments and is entrained by the ram pressure of the radio bubble, which itself is thought to be produced by 
cosmic-ray driven outflows at the Galactic center.  In this interpretation, the gas streamers on the western side of G0.17+0.15 are stripped, 
accelerated from 0 to $\delta v\sim\,35$ \kms\, over a time scale 
roughly $8\times10^4$ years, implying that ablating ram pressure is $\sim700\, \mathrm{eV\, cm^{-3}}$, comparable to the $\sim10^3\, \mathrm{eV 
cm^{-3}}$  cosmic-ray driven wind pressure in the Galactic center region.
\end{abstract}


 




\section{Introduction} 

Energetic processes have produced the large-scale Fermi $\gamma$-ray bubbles toward the nucleus of the Galaxy 
\citep{su10}. The origin of this kpc-scale structure is 
thought to be a nuclear wind driven by  either nuclear activity 
or star formation \citep{crocker11,yang13}. Nuclear winds are well recognized in 
many star-forming galaxies \citep{veilleux05}, and the evidence for this hypothesis comes from high velocity 
clouds 
exceeding  $|v_{\rm r} | \sim 300$ km s$^{-1}$
at high latitudes entrained in the Fermi bubble \citep{mcclure13,diteodoro20}. Lower velocity $\sim$150 \kms\, HI 
clouds entrained in the Fermi bubble 
have also been detected closer to the Galactic center suggesting increasingly  higher velocities at 
higher latitudes \citep{diteodoro20,lockman20}.

Bipolar X-ray emission filling a 400-pc radio bubble emerging from the Galactic center has been detected 
\citep{ponti19,heywood19}. 
The question is whether a large-scale nuclear 
wind is also responsible for its origin. 
In this case, high cosmic ray pressure is thought to be driving a large-scale wind 
\citep{heywood19,zadeh19}. The cosmic 
ray pressure in the Galactic center region is much higher than the gas pressure \citep{oka05,zadeh19}, 
leading to a scenario in which the gas is dragged 
along by escaping 
cosmic rays  and  forms a nuclear wind 
\citep{kulsrud71,everett10,zweibel17}. Unlike the Fermi bubble, however, there is no 
evidence of  high-velocity entrained clouds associated with the nuclear wind producing the X-ray-filled 
radio bubble.
Radio recombination line (RRL) 
observations of warm ionized gas in the  eastern and western, high Galactic latitudes,   Lobes, 
identified at the edges of the radio bubble \citep{sofue84,tsuboi86}, 
 show low velocities ranging
between $\sim$20 and -20 \kms, respectively \citep{alves15}. These low velocities  
are inconsistent with the expectation that  the Lobes are  byproducts of the nuclear wind, 
suggesting that the Lobes may not lie in the Galactic center \citep{tsuboi20}.

Although the radio bubble could  result of a Galactic center nuclear wind, 
it  is important to 
identify high velocity clouds that are interacting 
with the underlying wind that expands  the bubble. We have identified an HII region, 
G0.17+0.15,  showing  a radial velocity of  
$\sim$130 \kms\, 
suggesting that G0.17+0.15 lies in the Galactic center.  
This source lies $\sim4'\, (\sim10$pc) toward high Galactic  latitudes (north)  of the  prominent thermal, 
ionized Arches  at  $-20$ \kms\,  \citep{pauls80,zadeh87,lang01} and appears to be located within  large-scale, high-velocity  $^{12}$CO molecular 
clouds traced in molecular line surveys of the Galactic center \citep{bally88,sofue95,oka98,henshaw16,sormani19,veena23}.



The  high-velocity, spectral 
index and polarization characteristics suggest that  G0.17+0.15 is a Galactic center object interacting 
with nonthermal radio filaments (NRFs). We report a strong helium radio recombination line emission from G0.17+0.15, likely be one of the 
strongest helium recombination  
line emission in the Galactic center region. 
The morphology and kinematics of  G0.17+0.15 
show  evidence of an external  flow   dragging  
and shredding   the ionized gas at the edges of the HII 
region.  
We argue  that the ram pressure of the   cosmic-ray 
driven wind inflating the radio bubble is able  to explain the velocity difference across the 
HII region and is sufficient to accelerate  the 
HII cloud.

In this paper, we explore one possibility that can explain the results of our measurements.   While each of the observational findings 
can be
explained in alternate ways, 
such as infall motion of the cloud and/or HII region from
its high-latitude location towards the Galactic center, 
the nuclear wind HII interpretation is interesting 
enough that we focus here in exploring its 
plausibility as 
opposed
to arguing that it is a unique interpretation. 
In particular, we  focus on G0.17+0.15 and present results of RRL measurements toward G0.17+0.15. 
This  cloud is  localized toward the eastern  Lobe, near $l\sim0.2^\circ$. 
We argue here that the G0.17+0.15 cloud is  embedded within the eastern Lobe, which is  dominated by a mixture of thermal and non-thermal radio 
emission,  
and that the filaments of the Radio Arc ($\sim0.2^\circ$) 
are interacting with a thermal HII region,  G0.17+0.15.  
We have also studied IR counterparts to the ionized gas and the energetics needed to produce  gas and dust 
emission.  
A true physical interaction will tell us about  the nature of cosmic-ray driven 
outflows  at the Galactic center and the possible resemblance of 
 the radio and Fermi  $\gamma$-ray bubbles.

\begin{figure}
\centering
\includegraphics[scale=0.2,angle=0]{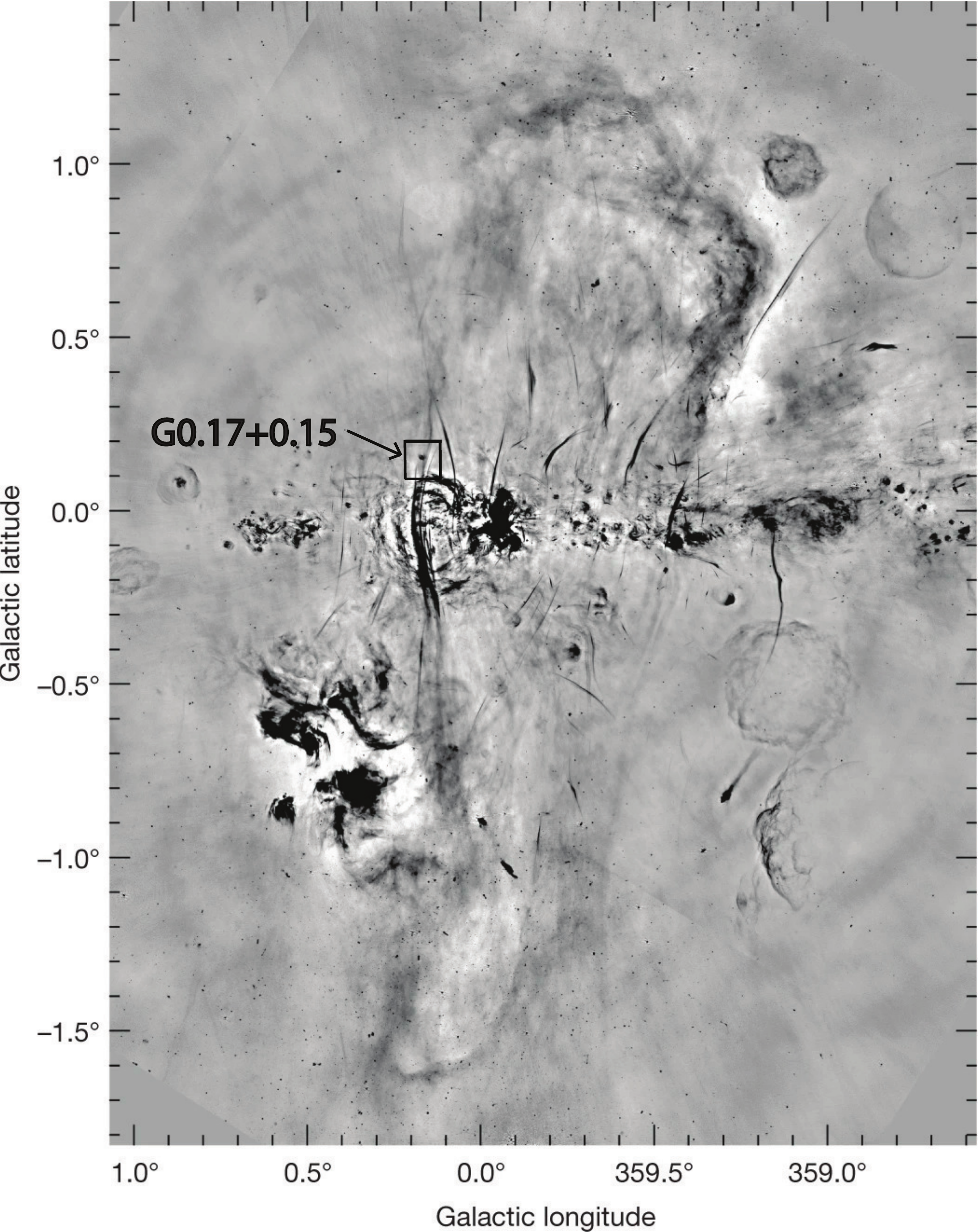}
\includegraphics[scale=0.7,angle=0]{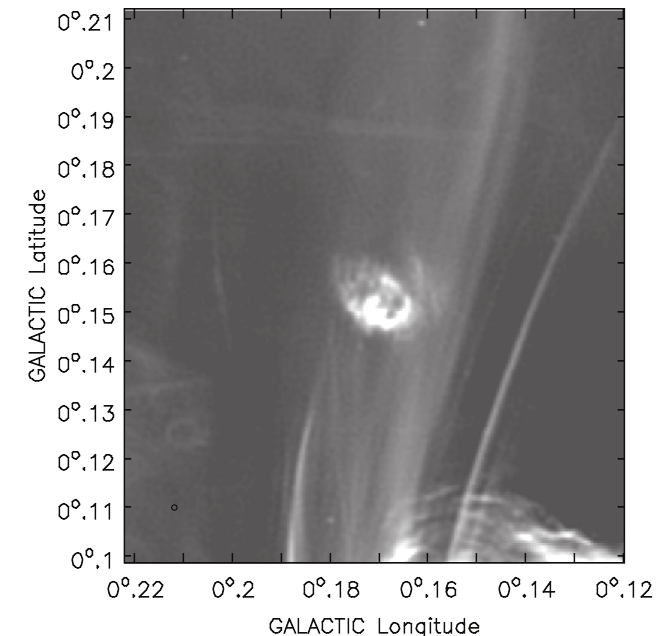}
\includegraphics[scale=0.7,angle=0]{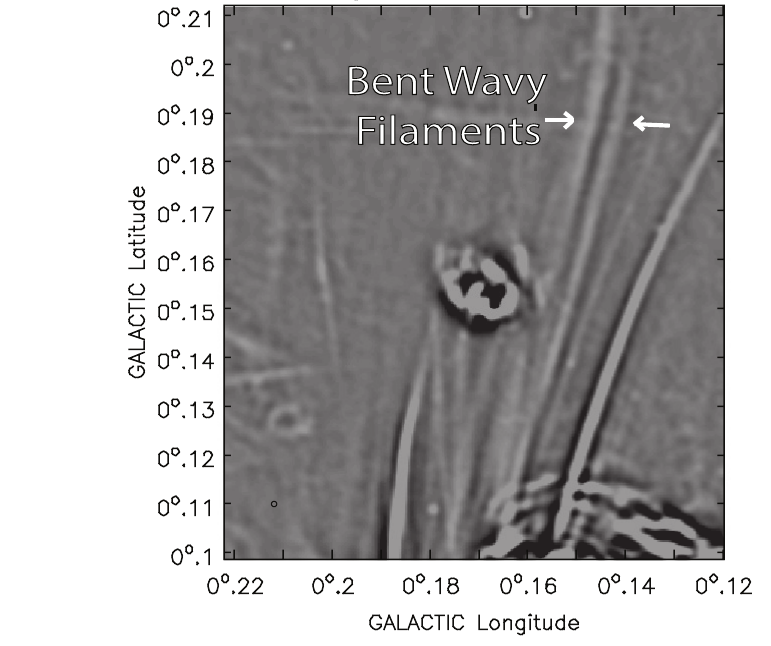}
\caption{
{\it (Top a)}
A 1.28 GHz mosaic of the radio bubble with a resolution of 4$''$, (Figure 1 of \citet{heywood19}).
 G0.17+0.15 is shown along the northern extension of 
the Radio Arc near $l\sim0.2^\circ$. A rectangular box  
shows the region surrounding G0.17+0.15. 
\citep{heywood19,zadeh22}.
{\it (Bottom Left b)} 
A  $0.1^\circ\times0.11^\circ$ 
grayscale  view of G0.17+0.15  showing multiple filaments to the west  of the 
HII region G0.17+0.15. 
{\it (Bottom Right c)} 
A filtered image of (b) showing three  supporting filaments disappear at three positions 
on the ring (see also the diagram in Fig. 2b).   
Starting from the south, the bent wavy filaments appear to be deflected to the west of G0.17+0.15 and the return to a northward direction after 
passing the region. Two arrow point to these filaments. 
The small open circles in (b) and (c) show the size of the synthesized beam using CASA.  
The field of view shown as a box in (a)  is displayed in (b) and (c). 
}
\end{figure}

\section{Observations and data reductions}

We used the GBT, MeerKAT and the VLA to  study the 
continuum emission, radio recombination line emission, and spectral index properties, as described below. 


\subsection{Continuum}
We have used 1.28 GHz radio continuum 
images of the Galactic center with 4$''$ and 6$''$ resolution based on unfiltered and filtered data taken 
with MeerKAT, respectively 
\citep{heywood22,zadeh22}. 
Details of filtered and unfiltered images  are described in \citep{heywood22,zadeh22}. 
We also used the VLA wideband data observed at  6 GHz   (C-band) and 10 GHz  (X-band). 


\subsubsection{6 GHz-band  VLA Data}

The C-band observations with the VLA were carried out at three epochs: 2022-06-28, 2022-07-03 and 2022-07-04 in the A-array
configurations with a total on-source time of 150 min, sampling the spectral polarization data of RR, RL, LR and 
LL with 64 channels of 2-MHz width  in each of 32 sub-bands. The total bandwidth of the continuum data is 4 GHz centered
at 6 GHz. The VLA data reduction  was  conducted using CASA \citep{casa22}. 
The high-resolution C-band data were initially calibrated using the VLA pipeline program for the gains, bandpass,
polarization leakages and phase offsets as well as flux-density and absolute
 linear polarization angle (the results of these measurements will be 
given in a future paper discussing compact radio sources in the Galactic center).
The residual errors were further corrected using better models, including an extended structure 
in addition to a point-like core of the calibrators J1733-1304 and J1744-3116 following the procedure described by \cite{zmg2019}. 
 Then, the HII complex and its surroundings
were imaged and cleaned using $tClean$, a CASA program, weighting the visibilities with a robustness parameter 
$R$=0.65 \citep{briggs95}. The rms noise of the final image  
is 3.5 $\mu$Jy beam$^{-1}$ convolved to a FWHM beam of $1''\times0.5''$ with position angle (PA) of 0$^\circ$.        
Polarization and in-band spectral index measurements will be given elsewhere. 
The phase center is at RA (J2000) = 17:45:26.400; Dec (J2000) = $-$28:42:46.500. 


\subsubsection{10 GHz-band  VLA Data}

The X-band observations were carried out in the D and C array configurations at seven epochs: 
2022-07-26, 2022-07-29, 2022-07-30, 2022-09-30, 2022-10-01, 2022-10-09 and 2022-10-16. The telescope 
system was set up for simultaneously observing the hydrogen recombination lines 
at the eleven 
Hn$\alpha (83< n < 92)$ 
in each of 11 sub-bands in the frequency range between 8 and 12 GHz.
Each sub-band consists of 1280 channels with a width of 0.05 MHz (a mean velocity width of $\sim1.5$ \kms).
 An effective bandwidth
for the continuum data is therefore $\sim$0.70 GHz. Corrected for residual errors, the X-band visibilities were used to
construct an image for the continuum  emission, achieving an rms noise of $\sim10 \mu$Jy beam$^{-1}$, convolved to a FWHM beam 
of 4.6$''\times2.1''$ (PA$\sim10^\circ$). 
The phase center is at RA (J2000) = 17:45:26.40; Dec (J2000) = $-$28:42:46.50. 


\subsection{Radio Recombination Line (RRL) Data}

\subsubsection{10 GHz-band  GBT Data}

RRL observations of G0.17+0.15 using the Green Bank Telescope (GBT) at 10 GHz were part of an (unpublished) targeted survey of Galactic center 
sources with the GBT in 2005.  The observations detected six Hn$\alpha$ (87 $\leq$ $n$ $\leq$ 92) and nine Hn$\beta$ (109 $\leq$ $n$ $\leq$ 117) 
transitions found within the X-band receiver's bandpass (8.0 - 10.0 GHz) over multiple IFs.  An On-Off observing technique with a 
roughly with $\sim70''$  beam was employed for calibration and the recombination lines were co-added in GBTIDL to improve sensitivity.  The total 
achieved velocity range exceeded 1000 km s$^{-1}$~with a velocity resolution of less than 1\kms~after hanning smoothing.  A sparsely sampled map was produced 
with 32 pointings towards G0.17+0.15 with a final spatial resolution of roughly 1.6\amin~(FWHM). 

\subsubsection{10 GHz-band  VLA Data}

We followed GBT observations with   high-resolution RRL observations of eleven  hydrogen 
recombination lines  Hn$\alpha$ (82 $\leq$ n $\leq$ 92) over a 4-GHz bandwidth of X-band  with the VLA, 
as described in section 2.1.2.
The corrections for the residual errors were done  on the basis of sub-bands 
using the gains and bandpass derived in
corrections for the continuum visibility data. The  corrected uv-data observed at the 
seven epochs were combined. 

The line free
channels were chosen to make linear fitting  to the continuum baselines, 
in the uv-domain across the 1280 spectral channels
for each of the eleven sub-bands. This was followed by 
removing  the continuum emission  using
the linear-fitting results. The fitting and
continuum subtraction were carried out by utilizing the CASA task UVCONTSUB. Then, 
the eleven spectral line cubes were transformed to 
the LSR velocity frame in the uv-data domain while setting up the rest frequency using the corresponding 
hydrogen $\alpha$-transitions. The spectral line visibilities were  regridded in the 
velocity coordinate with velocity width of 2 km s$^{-1}$. The eleven hydrogen $\alpha-$line image cubes 
were constructed using $tClean$ in a  velocity  range from $-200$ to 400  km s$^{-1}$,
and the dirty image cubes were cleaned using the Hogbom algorithm \citep{hogbom74}, achieving in constructing and cleaning 
a typical rms noise in the range
between  0.25 and 0.5 mJy beam$^{-1}$ depending on the noises of sub-band data. We also noted that a linear function used to 
interpolate the robustness weight parameter $R$ across the eleven sub-bands so that
the baseline visibilities of eleven transitions are properly weighted to produce a synthesized beam
equivalent to that of $R$=2 for  the H82$\alpha$ line image.   
In the Appendix \S7.1, we describe how the RRL image cubes of eleven RRLs were stacked in order to enhance the signal-to-noise ratio.

\subsection {In-band Spectral Index}

Details of MeerKAT observations,  data  reductions and images (unfiltered) are presented  in \citep{heywood22}.  
Using  a filtering technique, filtered MeerKAT images  are presented in \citep{zadeh22}. 
In-band spectral  indices 
are determined from 
broad bandwidth 20 cm data  
using  unfiltered and filtered MeerKAT data with a resolution of 6.4$''$  and 
8$''$,  respectively \citep{heywood22,zadeh22}. 
\subsubsection{VLA Data}
Details of in-band spectral index using VLA wideband data can be found in the 
Appendix \S7.2. We present below spectral indices of  unfiltered and filtered  images.  







\label{lastpage}

\section{Results}

Here we mainly focus on the physical relationship between nonthermal radio filaments bundled together by  the   Galactic center Lobe (GCL) and 
the HII region G0.17+0.15, located in  the eastern side of the high-Galactic latitudes  GCL. 
 
\subsection{G0.17+0.15}

Multiple  morphological arguments given below,  favor  an interaction between  G0.17+0.15, 
resembling a  ring with clumpy structure, and the NRFs of the Radio Arc. 
However,
 morphological arguments
alone tend to be inconclusive, thus,  we 
will show structural details of G0.17+0.15 aided by kinematics, 
spectral index  and RM measurements to argue that  G0.17+0.15 is a candidate 
interaction site. 
this remarkable high 
velocity ionized 
cloud G0.17+0.15  shows similar characteristics to  high-velocity clouds  associated with the   $\gamma$-ray bubbles.


\begin{figure}
\centering
\includegraphics[scale=0.8,angle=0]{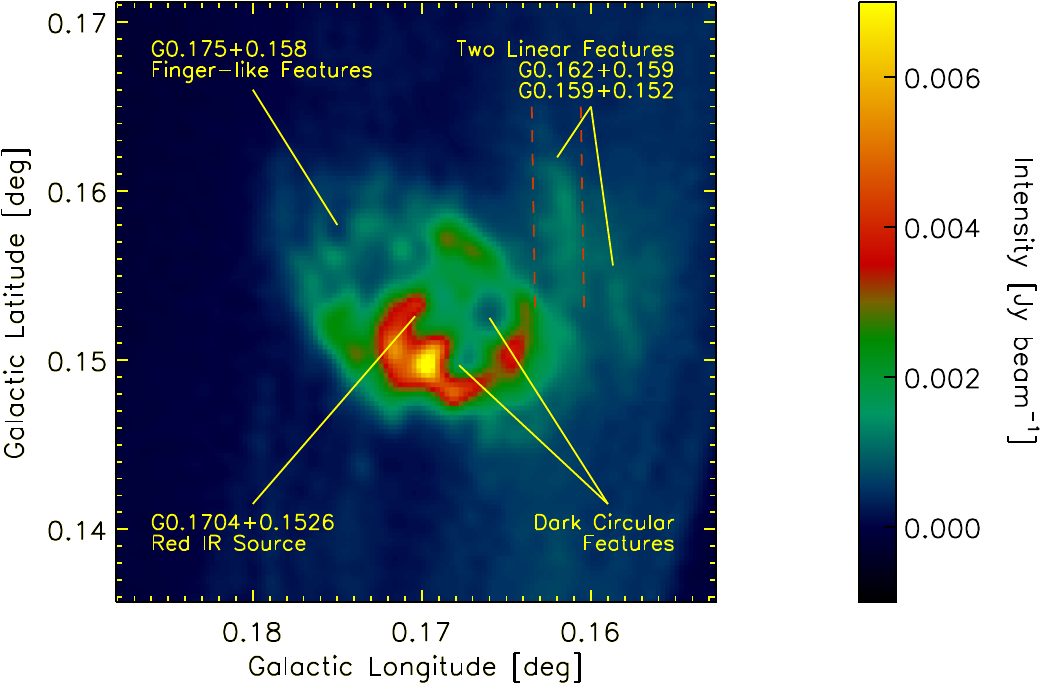}
\includegraphics[scale=0.26,angle=0]{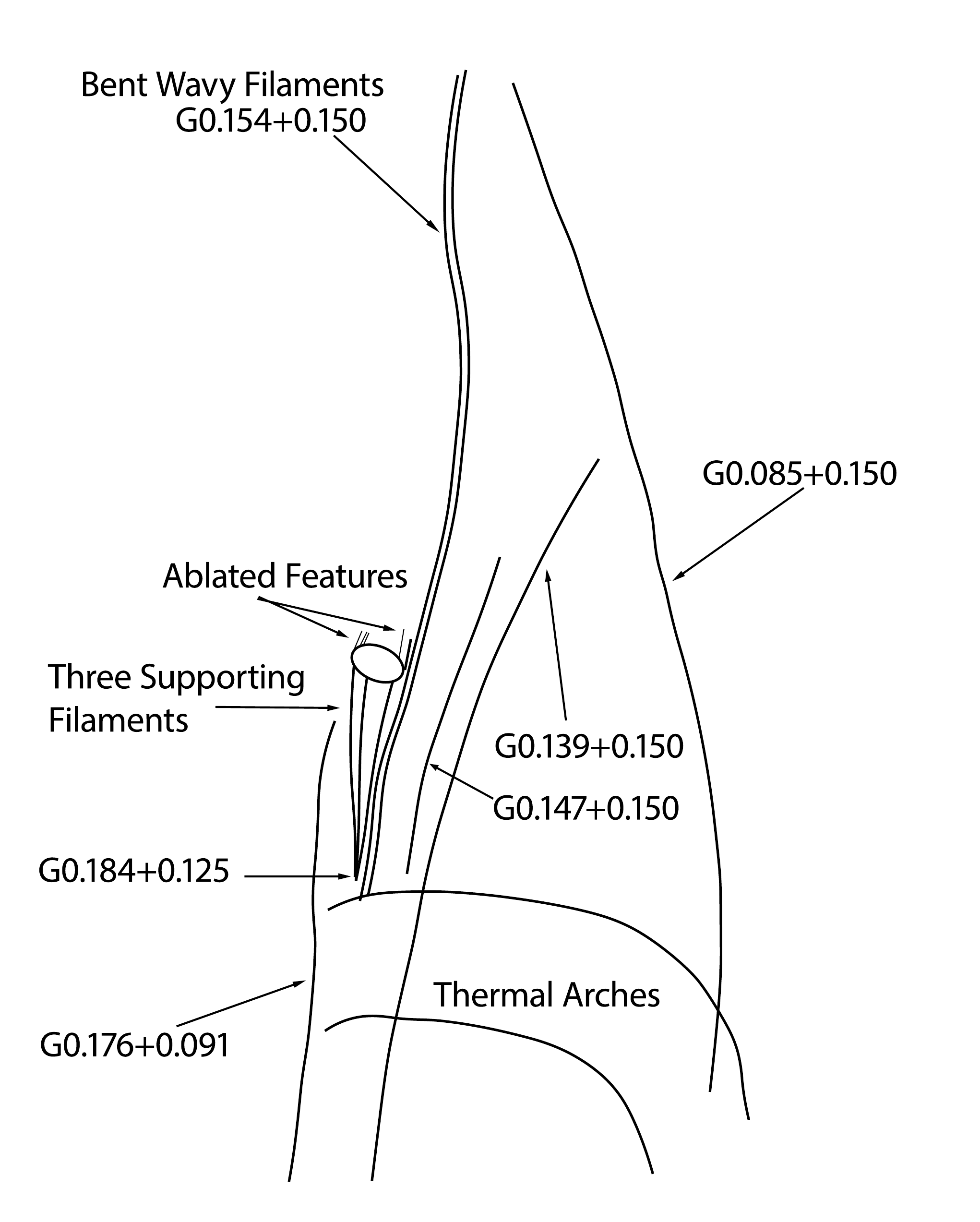}
\caption{
{\it (Top a)}
A close-up continuum view of G0.17+0.15 at 1.28 GHz with a resolution of $0.4''$  showing its clumpy ring-like  
envelope and a  bow-shock structure within the ring.  Major features in the HII region are 
labeled.  The color bar units are in Jy beam$^{-1}$. The parallel lines (red dashed lines)
indicate the region from which position-velocity image is 
constructed, as shown in Figure  7. 
{\it (Bottom b)}
A schematic diagram showing the  features associated with G0.17+0.15 and its surrounding filaments. 
}
\end{figure}

\subsubsection{Morphology}

Figure 1a shows a degree-scale panoramic 
view of the radio bubble tracing the edge of the northern GCL and the radio bubble  at 1.28 GHz \citep{heywood19}. 
In this MeerKAT image, we note G0.17+0.15 and nonthermal filaments as well as the thermal Arches to 
the south. 
This HII region,  G0.17+0.15 with its ring-like  
 appearance,  lies along the eastern GCL,  and appears to be embedded within  a large number of vertical nonthermal filaments.

Figure 1b,c  show 
unfiltered  and filtered close-up views of 
the HII region over a $\sim6'\times6.6'$ field at 1.28 GHz. 
The HII region   resembles   a clumpy, ring-like  envelope with  a 
 bowshock exterior  to the ring (see also Fig. 3 of \citep{zadeh88}).  
We note a gently bent bundle of filaments with their  wavy structure  
along the western edge of 
G0.17+0.15 extending to high latitudes $b>0.28^\circ$. 
There are also  other  isolated nonthermal filaments noted to the 
west of the ring that do not appear  to be related to 
G0.17+0.15. These filaments 
to the west of the ring at constant latitudes are  
 G0.147+0.15, and G0.139+0.15 roughly 1.4$'$ and  
2$'$ from the ring, respectively,  gently bend but are not distorted and do not have  a wavy 
appearance (see Figs. 1b,c). Figure 2a shows a color image with major features are labeled and Figure 2b shows a schematic diagram showing 
structural details discussed here.


Using the VLA, another rendition,  a G0.17+0.15 
with a resolution of  $4.6''\times2.1''$ at 10 GHz is shown 
in Figure 3a. 
A close-up  view of this image in Figure 3b shows 
clumpy distribution of the ring, with a southern and northern shell-like structures   
and two linear features  with an extent of 
$\sim0.6'$ running vertically  near  $l=0.162^\circ$. The brightest continuum emission is at 
G0.170+0.149 which lies at the southern shell  of the ring. 
The  disjoint linear 
features give   the appearance that they are connected to the arc-like structure at the SW of the ring (see Figs. 2a, 3b). 
We also note two dark  circular spots  with a diameter of $\sim12''$ each 
centered at G0.166+0.168, 
G0.152+0.149,  respectively,  in the   
western half of the ring.  These dark spots  may trace dense molecular gas 
or a cavity in the ionized gas  \citep{zadeh12}.
The southern-half of the ring  is sharper in the south   
than in the north, giving the appearance of a bow-shock structure.

Based on the 1.28 and 10 GHz continuum images, we make morphological arguments that there is interaction between 
the filaments and G0.17+0.15. 
First,  the wavy pattern or meandering  of the 
filaments at the western edge of the ring (see Fig. 1c) could  be due to 
an interaction. 
In this picture, the distortion of the filaments results from the encounter of the filaments with 
G0.17+0.15. Second, the western edge of the ring is  brighter 
than the eastern side by a factor of 3-4.  The two linear features near G0.162+0.159 and G0.159+0.152 with an extent of 0.6$'$  
are disjoint from the ring. The linear features could be produced by the ionized gas at the western edge of the ring being stripped and dragged along 
the direction of the bundle of nonthermal radio filaments (NRFs). In \S3.1.3, the kinematics of these linear features are described.
Third, we note  that the 
eastern half  of the ring  shows five short, parallel finger-like features running perpendicular to the ring with typical lengths of 
$\sim18''$ (see Figs. 1b,c)  and  
surface brightness  $\sim0.5-0.8$ mJy beam$^{-1}$  at 1.28 GHz.  
These finger-like features  near G0.176+0.158 
are oriented along the 
vertical filaments of the Radio Arc suggesting that 
ring material is stripped off by an external pressure and  dragged along 
the  magnetic field lines of NRFs. 
The finger-like and two linear features are suggestive of  ablated features (see Fig. 2b). 
Fourth, several NRFs appear to arise from thermal  Arches near G0.176+0.091 
and disappear at the southern edge of the ring at three locations 
(G0.167+0.147, G0.172+0.148 and G0.177+0.091), giving the appearance that they are 
supporting  the ring from the south (see supporting filaments in Fig. 2).  
Multiple  filaments with diffuse emission at the westernmost  edge 
of the ring G0.154+0.150 bend with a wave-like structure  
along their higher Galactic latitude  extension.  
Meandering of the total and polarized intensities of NRFs  have also 
argued as a possible sign of an interaction \citep{zadeh88}.
A gentle wave-like bending of NRFs  at the 
western edge of the ring,  
a north-south bow-shock appearance of the ring, termination of the filaments at the southern edge of the 
ring  and 
finger-like vertical structure on the eastern side of the ring 
all suggest  the impact of an external source of pressure, directed from the south,  
impacting  the HII region (see Fig. 2b). 

\begin{figure}
\centering
\includegraphics[scale=0.6,angle=0]{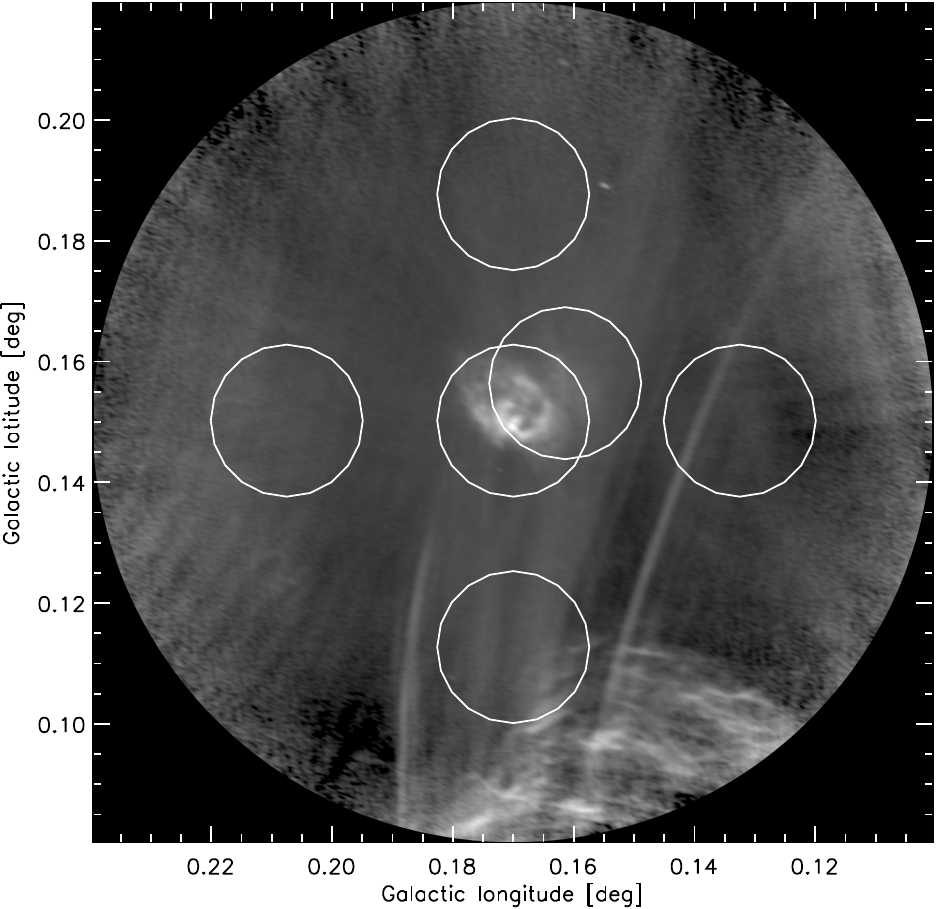}
\centering
\includegraphics[scale=0.6,angle=0]{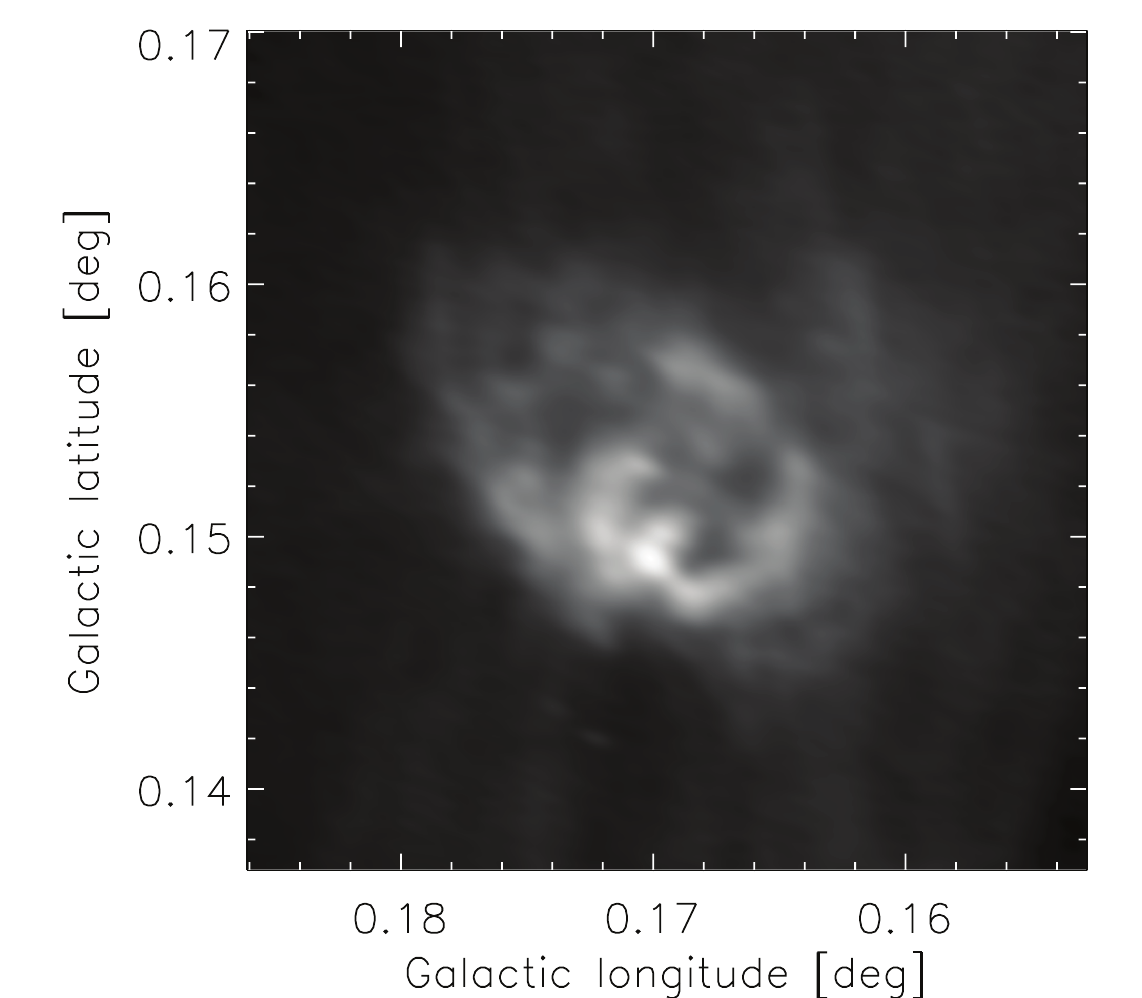}
\caption{
{\it (Top a)}
VLA image of G0.17+0.15 at 10 GHz.
The image is convolved with a beam of FWHM = $4.6''\times2.1''$ (PA=$68.6^\circ$).
The phase center is $l = 0.170^\circ, b = 0.150^\circ$ 
or  RA (J2000) = 17:45:26.40; Dec, (J2000) = $-$28:42:46.5.  
The rms noise $\sigma_{\rm rms}$ is  $10~\mu$Jy beam$^{-1}$.  
Circles drwan on the figure indicate the positions from which 6 spectra (see Fig. 4)  were extracted. 
Circles  The size of a circle represents the GBT beam FWHM=1.6$’$.
{\it (Bottom b)}
A close-up view of (a) showing clumpy structure as well  as as an arc-like halo 
structure that surrounds the  southwestern edge of the ring.  
}
\end{figure}

We also note that the size of the ring is similar to  the span  of the bundle of filaments that are 
grouped together, suggesting  that G0.17+0.15  is 
embedded within a large number of parallel filaments. 
An east-west 
asymmetry in the brightness distribution of NRFs crossing the HII region, is similar to that of the ring itself, 
suggesting that the two 
are related. As such, the western edge of the ring and NRFs to the west of the ring show an imprint of an 
interaction.



To  summarize,   a number of morphological arguments 
suggest  that the nonthermal filaments are  physically interacting with G0.17+0.15:\\

\noindent$\bullet$
Three  vertical  filaments terminate    at the location of the ring. This 
suggests that the filaments are interacting with the ring from the south  and 
are prevented from continuing to the north.\\ 

\noindent$\bullet$
The western half of G0.17+0.15 is brighter than the eastern half. Similarly, the large-scale 
bundle of filaments is   also brighter on the   western edge of the ring.\\

\noindent$\bullet$
Isolated, short,  vertical features Northwest of 
the ring G0.17+0.15 suggest that they are dragged away from the ring by the bundle of filaments. 
Similarly,  vertical fingers of ionized gas in the eastern half of the HII region  suggest 
 that they are likely  
stripped and dragged along 
the magnetic field direction of NRFs.\\ 


\noindent$\bullet$
The bundle of nonthermal  filaments 
along the western edge of G0.17+0.15 are distorted displaying a wavy pattern with a wavelength 
of $\sim10'$ (24 pc). This pattern is not detected  along 
the filaments that do not intersect with G0.17+0.15.\\


\noindent
Morphological arguments are always difficult to confirm. However,  kinematic data provide 
additional support for the interaction picture.  We first present low-spatial resolution recombination line data taken with the GBT,  
followed by higher resolution data taken with the  VLA. 

\subsubsection{GBT Kinematics: detection of high velocity ionized gas}

\begin{figure}
\centering
\includegraphics[scale=0.6,angle=0]{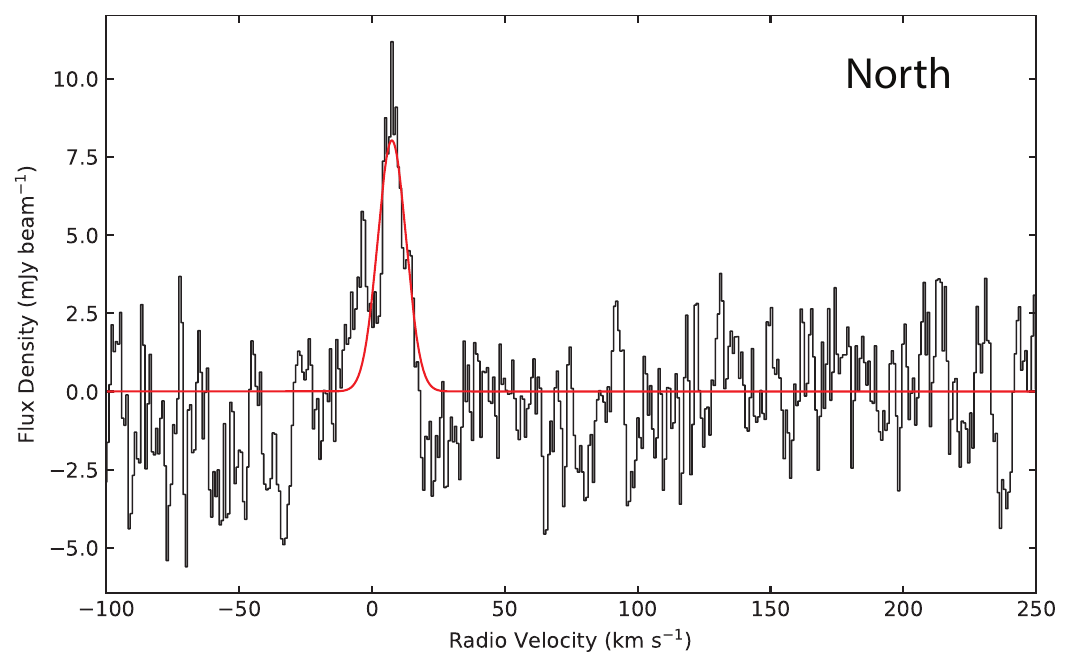}
\includegraphics[scale=0.6,angle=0]{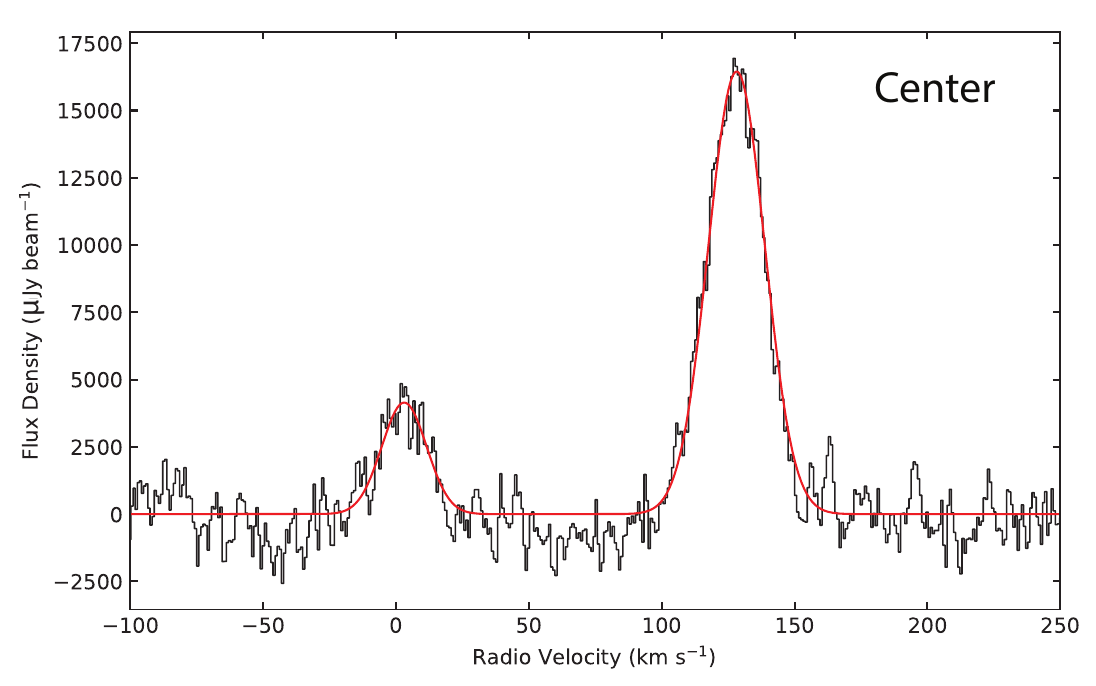}
\includegraphics[scale=0.6,angle=0]{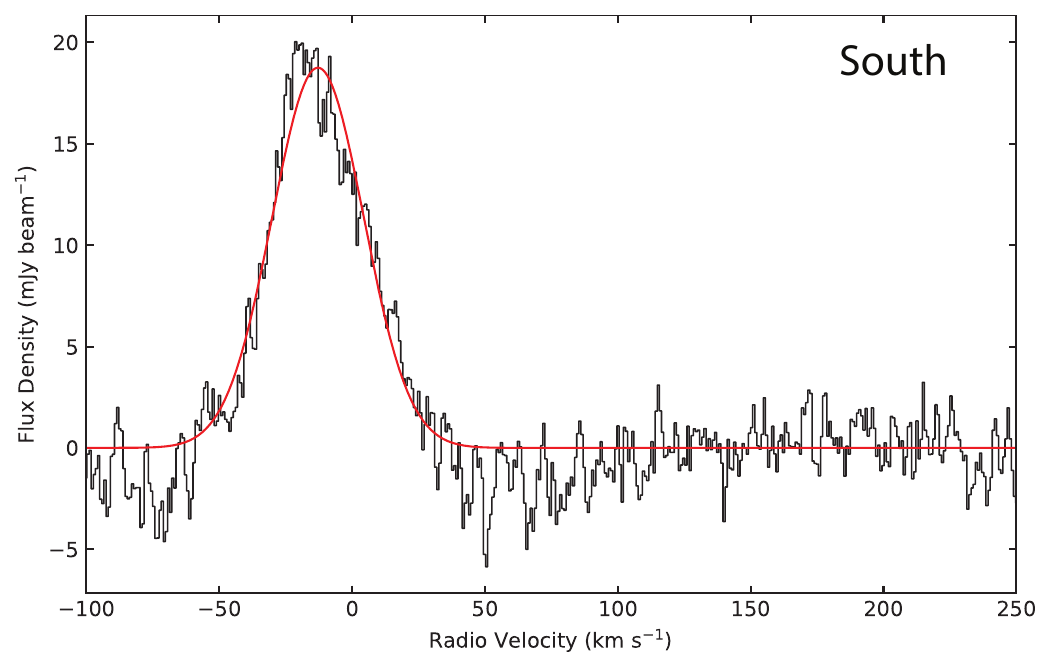}
\caption{
{\it (a)}  
Spectra of H$\alpha$  RRL emission per one 1.6$'$ beam from   3 positions 
(Center, North, and South) 
covering the HII region. 
Red curves show Gaussian fits to 
each spectrum and  fit parameters  can be found in Table 1 (the units in the center spectrum is in $\mu$Jy beam$^{-1}$ and all others in mJy beam$^{-1}$). 
The positions from Table 1 are marked
as circles on Figure 3a. The size of the circle corresponds to the beam.}
\end{figure}

\addtocounter{figure}{-1}
\begin{figure}[h]
\center
\includegraphics[scale=0.6,angle=0]{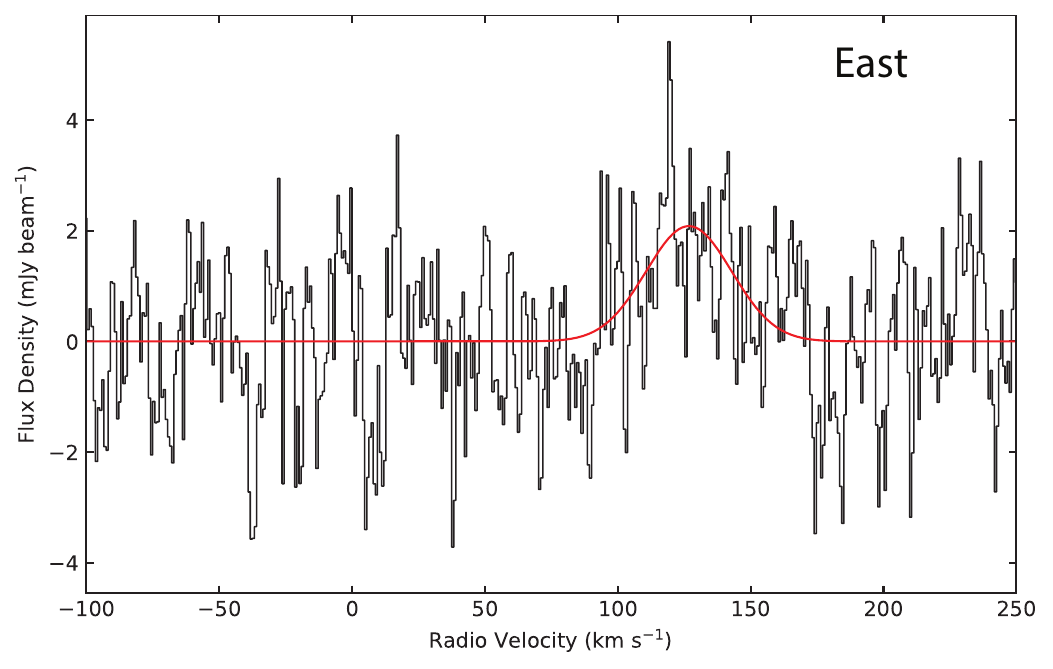}
\includegraphics[scale=0.6,angle=0]{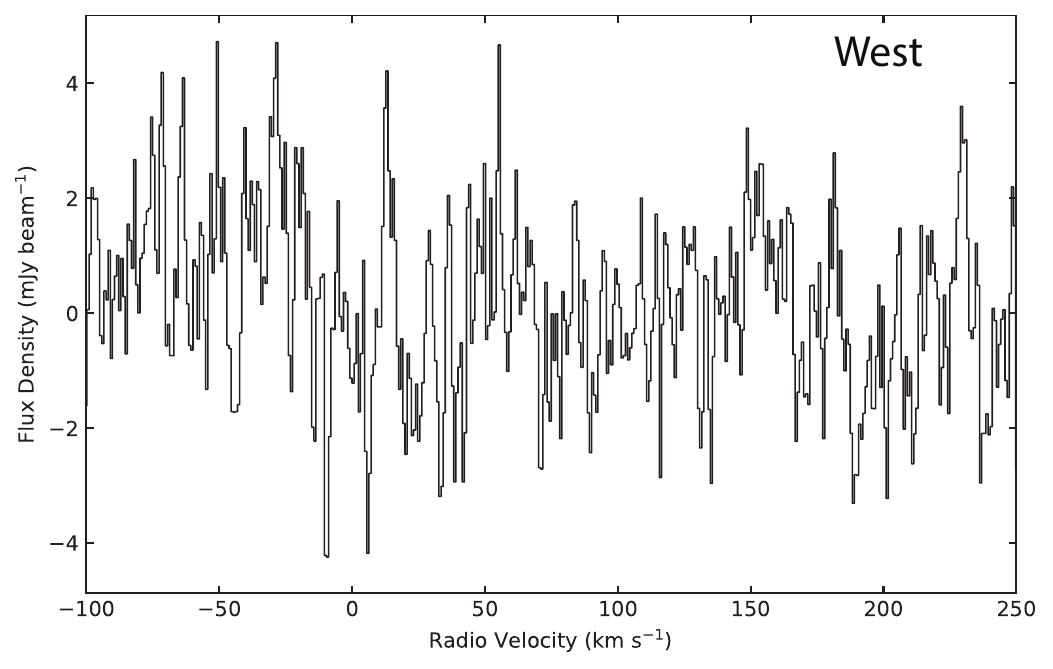}
\includegraphics[scale=0.6,angle=0]{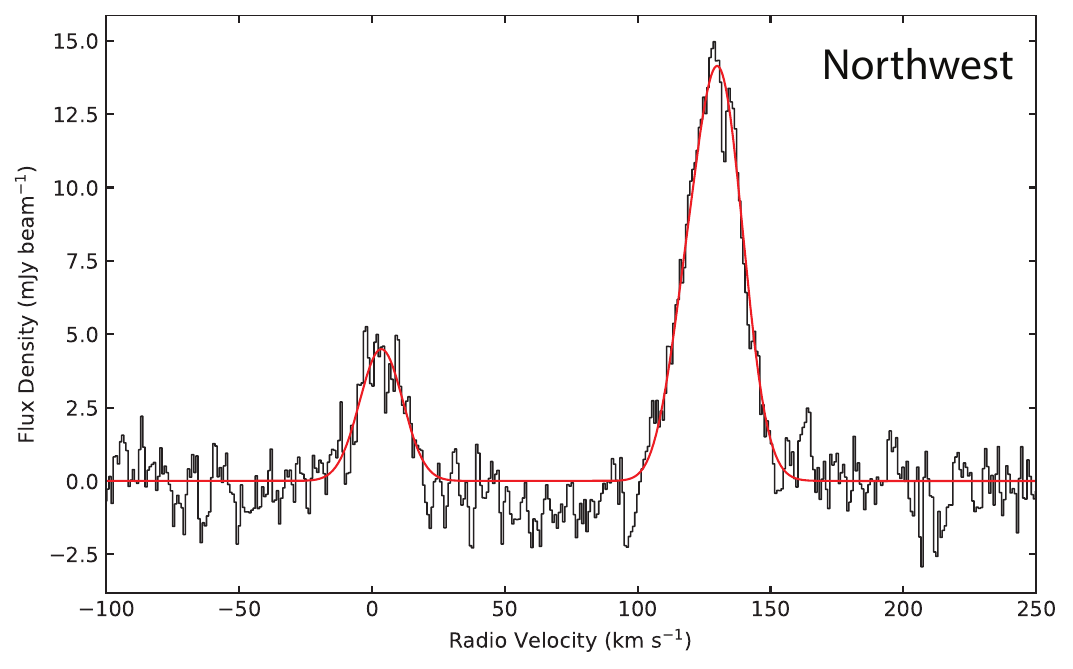}
\caption{
{\it (b)}  
Same as (a)  except for 3 different positions (East, and  West, and Northwest). 
}
\end{figure}

Figure 4a,b show six  spectra taken with the GBT toward  G0.17+0.15,  each separated  by one 86.4$''$ beam  with the exception of the  
Northwest  spectrum  
being 
offset from the center by a fraction of the beam width. 
The fitted spectra  show a high-velocity $\sim128$ \kms\, component,  localized to the ionized ring,  and a
low-velocity components, $\sim3.1$, 7 and -12 \kms components, which are  discussed below. This relatively high velocity ionized gas was 
first reported by \citep{2011ASPC..439...31R}. 
Columns 1 to 6  of Table 1 show the location  of each spectrum, and the  parameters of Gaussian fits: the center velocity, 
peak flux density  integrated over one beam,  
and  the linewidth. 
The high intensity ratio 
of the low-velocity  features  with respect to high-velocity component ($\sim30\%$) in the Center and Northwest positions  
suggests that
the spectral features within the large GBT beam,  near 
zero  \kms are  plausibly the helium recombination line mixed in with other lines from heavier species, such as CII. 
These lines are found in a larger, cooler and photo-dissociation regions 
surrounding the HII region as well as in the extended Galactic 
background and foreground RRLs throughout the Galactic center region (M. Royster et al.  2024, in preparation).
If a slightly bigger region were  selected, the 0 km/s diffuse gas would have been detected.
The 0 km/s diffuse gas has also been detected in $R(1,1)$ line of $H_3^+$ absorption
throughout the CMZ and  has been argued to  lie in the  Galactic  center region \citep{oka05,oka20}


We note that  the spectrum centered at G0.161+0.156 (Northwest position  on Fig. 4)  
shows  two 
adjacent  high-velocity components to the Northwest position of G0.17+0.15.  
The highest velocity component disappears to the east of the HII region. 
We note an east-west asymmetry in the radial velocity of ionized gas in the ring.  The 
highest velocity $\sim132$ \kms\, component is detected to the Northwest  of the ring when compared to 
the eastern spectrum with a velocity of 127 \kms. A larger velocity difference is noted in high-resolution VLA data discussed in \S3.1.3.

Another interesting asymmetry in the distribution of ionized gas outside the ring is noted in the six  spectra 
displayed  in Figure 4a,b. 
The low  velocity component near  0  \kms\, runs in the center  of the HII region and  along the north-south direction outside G0.17+0.15. The 
linewidth of the 
South spectrum is broad because of contamination by the negative velocity ionized gas associated with 
the thermal Arches.
The spectrum of the position centered on  
G0.17+0.15, located  $\sim4'$ north of the 
negative velocity $-20$ \kms\, ionized gas cloud associated with the thermal Arches,  is shown in Figure 4a,b.  
 The nature and location of the pervasive low-velocity component in the Galactic center is 
not established \citep{law09,alves15,oka22}, but the limited RRL available data suggests a possible association 
between the two components. In this picture, 
the  low-velocity component is produced by weak population inversion. 
The continuum flux for the central pointing is 0.5 Jy beam$^{-1}$, implying a line-to-continuum ratio of 0.8\%. 
The continuum flux for the east, west and south  pointings are 0.06, 0.02 and 0.2 Jy beam$^{-1}$, suggesting
peak RRL fluxes of 0.5, 0.2, and 1.6 mJy beam$^{-1}$, respectively. The predicted levels fall short of detection 
limits for the east and west pointings, while the predicted RRL component for the north pointing is swamped  by 
the contribution from  the thermal Arches.

The low-velocity component of the Center  spectrum
could be due to helium recombination line 
emission which occurs at frequencies that are shifted by -122 \kms relative to their hydrogen counterparts, 
This would be an unfortunate coincidence of a low-velocity pervasive component and helium line in   G0.17+0.15. 
To test this, we used the VLA to localize the low-velocity component toward the HII region. 

\subsubsection{VLA Kinematics: high velocity gas,  and detection of  Helium recombination lines}

Figure 5a shows the  H$n\alpha$  line intensity in the channel image at 126 km s$^{-1}$  with  the peak of the hydrogen RRL having  S/N ratio 
$\sim50$. 
A velocity  profile at the peak emission 
is shown in Figure 5b where  we report  
a $>5\, \sigma$  helium line detection.
The reason that this line is unlikely to be due to hydrogen recombination line emission  is that that the emission is localized to the brightest 
region of the HII region. The zero \kms\, hydrogen RRL is extended and is seen throughout the Galactic center.   
Contours of the helium line emission (black) are superimposed on Figure 5a, suggesting that the integrated helium line intensity 
appears to be 
extended along the southern edge of  G0.17+0.15, peaking at  G0.17172+0.15105. 
This emission is localized only to the brightest region of G0.17+0.15, unlike the diffuse, pervasive   zero \kms\, feature that is 
seen throughout the Galactic center \cite{law09}. 


\begin{figure}
\includegraphics[scale=0.2, angle=0]{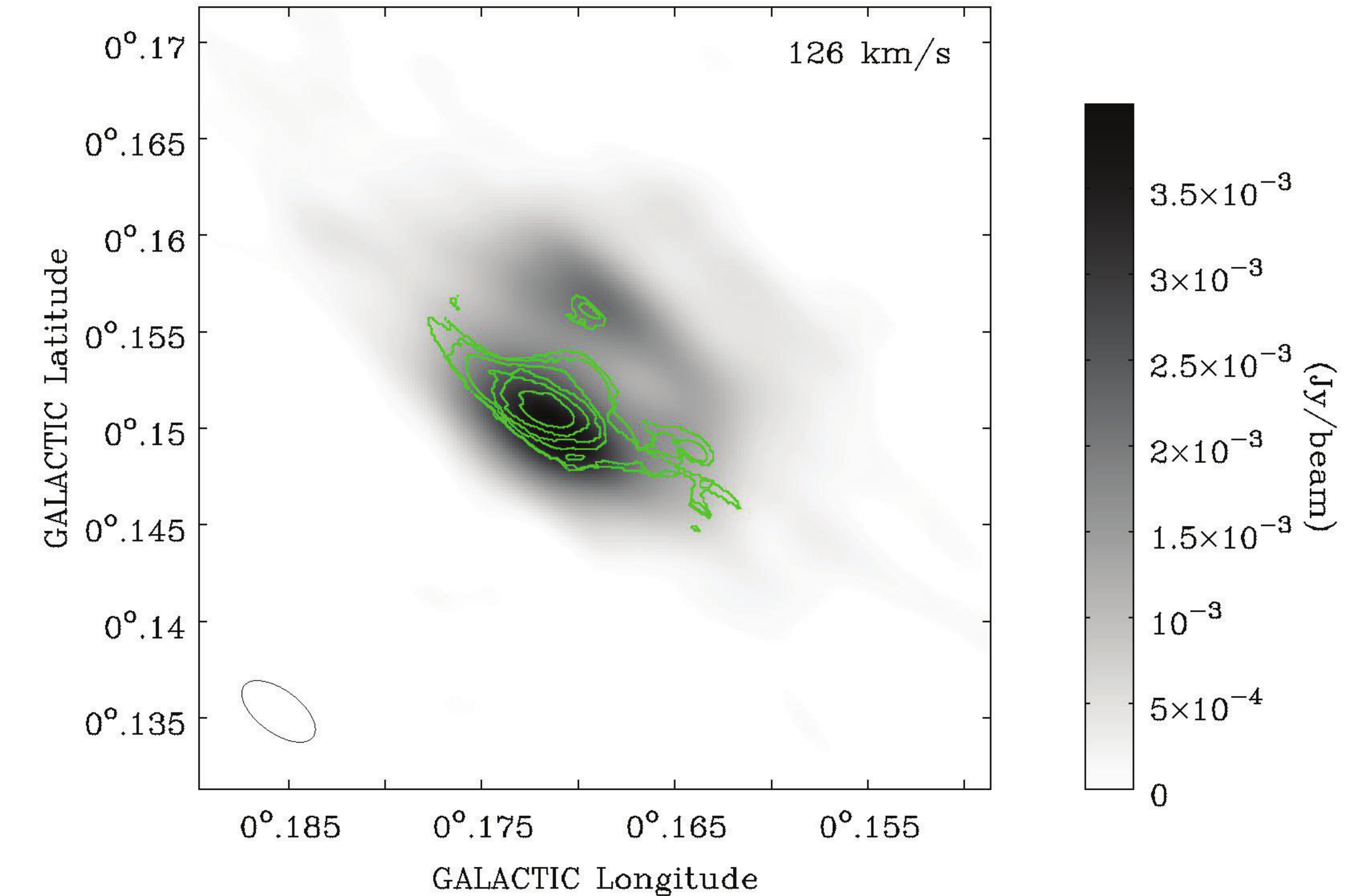}
\includegraphics[scale=0.35, angle=0]{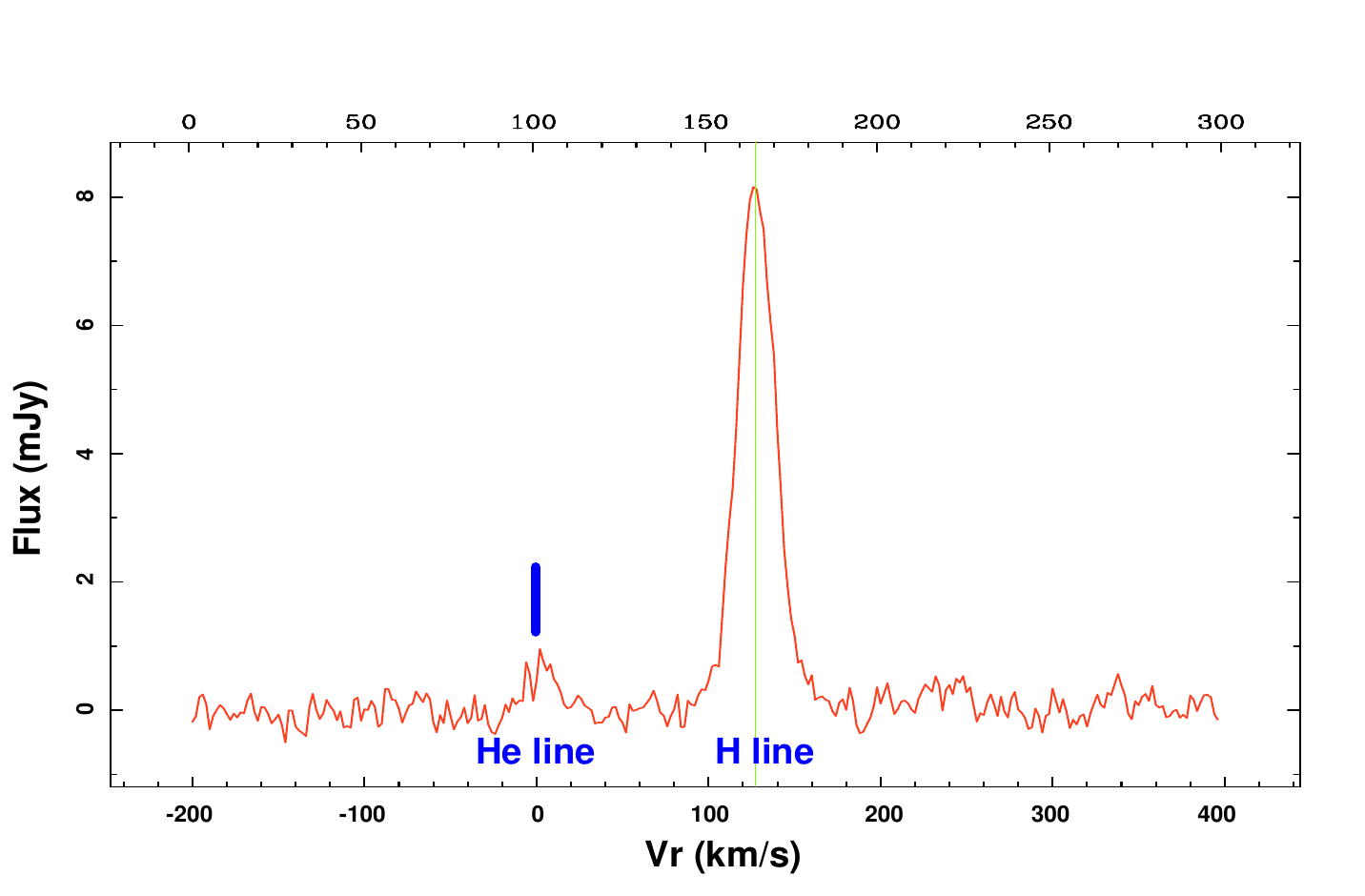}
\caption{
{\it (Left a)}
The channel image with the peak intensity of RRL at 126 km s$^{-1}$.
The contours (green) represent the intensity of  helium RRL flux 0.5 mJy beam$^{-1}$ km s$^{-1} \times$ (3, 4, 6, 8, 10 and 12). 
An open ellipse is shown at the  bottom-left corner 
corresponding to the synthesized beam $16''\times8''$ (PA=$53.5^\circ$). 
{\it (Right b)}
The line profile towards the region with the line peak. The hydrogen line peak is 
located  at 126 km s$^{-1}$, marked with a green  vertical line). 
The  corresponding Helium line is detected at near 0 km s$^{-1}$ for the hydrogen line but not for the helium line, marked with a blue bar. 
With a 4-$\sigma$ cutoff, we integrated the helium line flux from the channels at velocities between $-18$ and +18 km s$^{-1}$, 
as shown in the contour image.  
The top axis is labeled with the channel serial numbers of the spectral cube.
} 
\end{figure}

Two Gaussians were  fitted to the line profile at the peak emission, giving integrated line flux
values of 15$\pm$3.0 mJy  km s$^{-1}$  and 200 $\pm$3 mJy  km s$^{-1}$
for the helium and hydrogen lines, respectively, with  effective FWHM Gaussian widths of 25.3$\pm0.5$ km s$^{-1}$  for both lines.
Thus,  a value of  $\sim$ 7\% is inferred for the ratio of singly ionized helium
line to the hydrogen line.  
This  implies that this spectral feature is likely from the helium 
recombination line within the HII region.

\begin{figure}
\centering
\includegraphics[scale=0.85, angle=0]{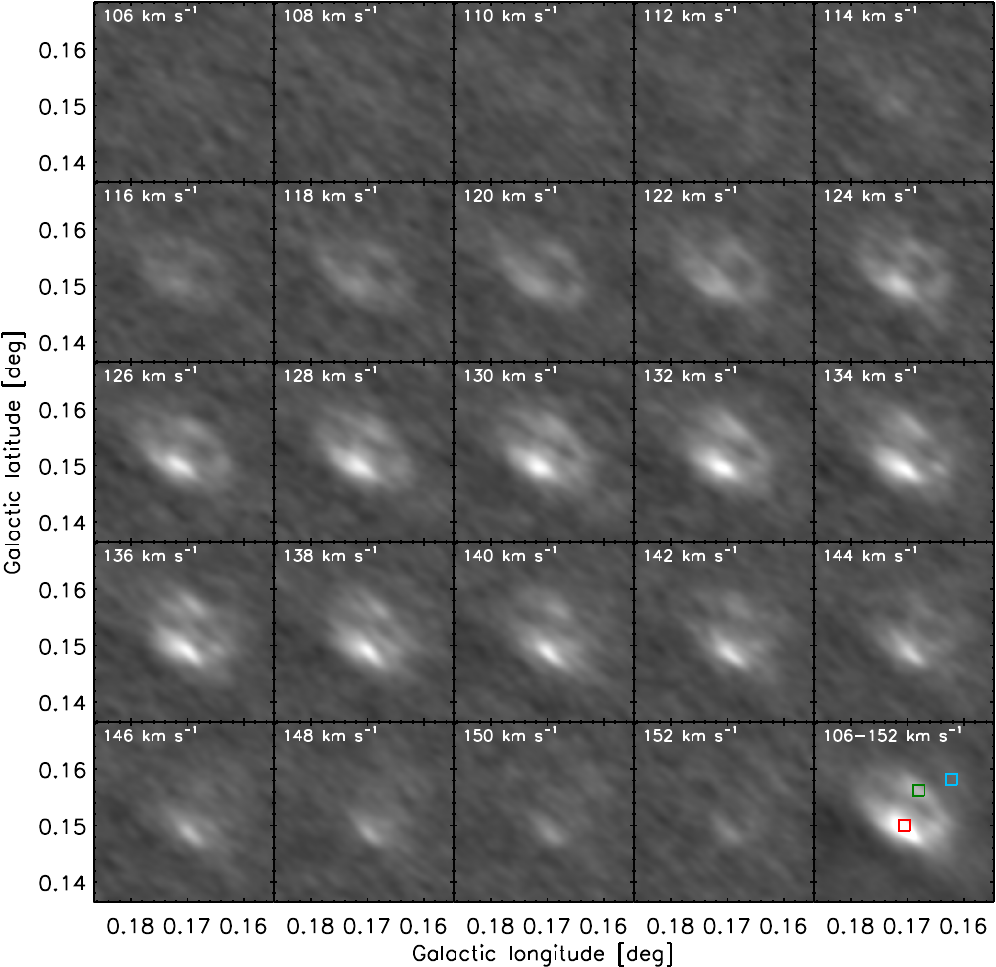}
\caption{
VLA image of the stacked RRL (H92$\alpha-$H82$\alpha$) from the HII region in G0.17+0.15 at 10 GHz, 
$\sigma_{\rm rms}$ = 65 $\mu$Jy beam$^{-1}$ ch$^{-1}$.
The image is convolved with a beam of FWHM = 16''$\times$8'' ($53.6^\circ$ in Galactic coordinates).
The LSR velocities of 100, 110, 130, 140, 150 km s$^{-1}$ are drawn on the left panels from top to bottom.
The velocity increment is $\Delta V=2$ km s$^{-1}$ from left to right. The integrated line intensity is shown in bottom right corner. 
The squares show the locations of three spectra displayed in Figure 7.
} 
\end{figure}

Figure 6 shows the results from the stacked RRL cube with a rms noise of 65 $\mu$Jy beam$^{-1}$.
The top panels show the 24 channel images at the LSR velocity from 106  to 152 km s$^{-1}$ among 
the 300 channel images in the RRL cube in a velocity range between $-$200 and 400 km s$^{-1}$. The 
bottom right image  shows the line emission integrated between 106 and 152 \kms.
The brightest line intensity is localized to the south of the ring near G0.170+0.149. The line profiles 
are broader in the southern shell.  

Figure 7 shows the velocity profiles of three   positions as indicated  on the  integrated line intensity image in the lower right corner 
 of Figure 6. 
The  velocity profiles of the northern  and  southern shells  of the ring
give $v_{LSR}\sim 126$ and $\sim129.5$ 
\kms.  The line emission is twice as strong in  the south than in the northern  shell of the ring. The Gaussian fitted linewidths of the 
southern and northern positions are 
26 and 22 \kms, respectively. 
The Gaussian fit to the third position centered on the faint linear feature shows 
a peak velocity of $\sim146$ \kms. 
This is the highest velocity of ionized gas detected toward the HII region. 
The position velocity diagram 
along  the  $\sim0.5'$ length of the linear feature (see Fig. 3b) is shown in Figure 7 (bottom right panel). 
 The velocity gradient along the disjoint linear feature G0.162+0.159 shows an increase to the north 
and is $\sim1$ \kms\, per  arcsecond or $\sim25$ \kms\, pc$^{-1}$. 

\begin{figure}
\centering
\includegraphics[scale=1.0,angle=0]{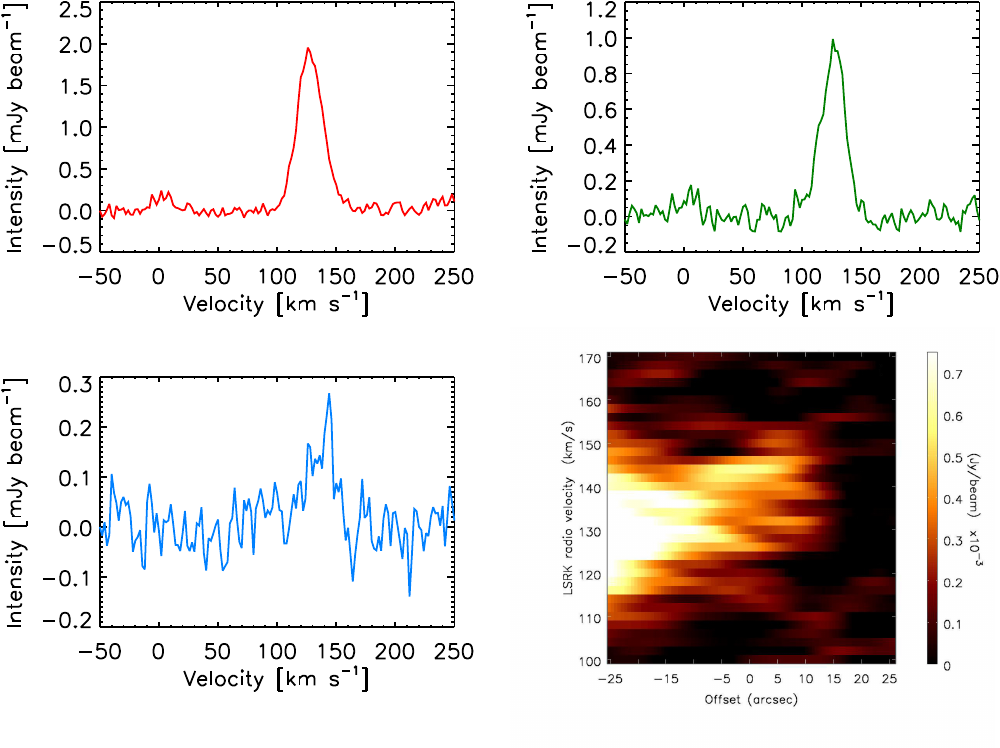}
\caption{The spectra of RRL emission from are presented toward three positions, as shown by squares  in Blue, Red and Green squares (see Text). 
The bottom right panels shows the position velocity (PV) diagram along the linear feature to the west of G0.17+0.15. The area from which the PV is 
 generated is shown by two parallel lines (see Fig. 2a) with  roughly a 
0.17$'$ and  0.5$'$ width and length, respectively, with PA = 180$^\circ$. 
The positive offsets  are  at higher latitudes than negative latitudes. 
The diagram covers the velocity range between 100  and  170 \kms. 
} 
\end{figure}


Numerous molecular surveys of the Galactic center show high-velocity gas in the general direction of G0.17+0.15
\citep{bally88,sofue95,oka98,henshaw16,sormani19,veena23}. 
Due  to the mismatch with  our high-resolution data, it is difficult to 
associate large-scale molecular gas with the HII region. 
We found no evidence of Far-IR emission 
at $\ge100$ $\mu$m that would be indicative of cold ($T \lesssim 20$ K) dust 
from the direction of G0.17+0.15 using  Herschel data.  
At 120-140 km/s, the nearest bright CO clumps are about $\sim0.1^\circ$  away. 
However,  there is  diffuse molecular gas in the direction of G0.17+0.15 at velocities between 120 and 140 \kms \citep{oka98}.  
In spite of a lack of cold dust emission at  the far-IR, the comparison of the CO molecular complex  and the HII region 
suggests that G0.17-0.15  could  be a star forming site  embedded within the CO cloud. 
Future higher spatial resolution  molecular line observations  would be capable of revealing a 
the possible association of the CO cloud complex with  G0.17-0.15.

\begin{figure}
\centering
\includegraphics[scale=0.3, angle=0]{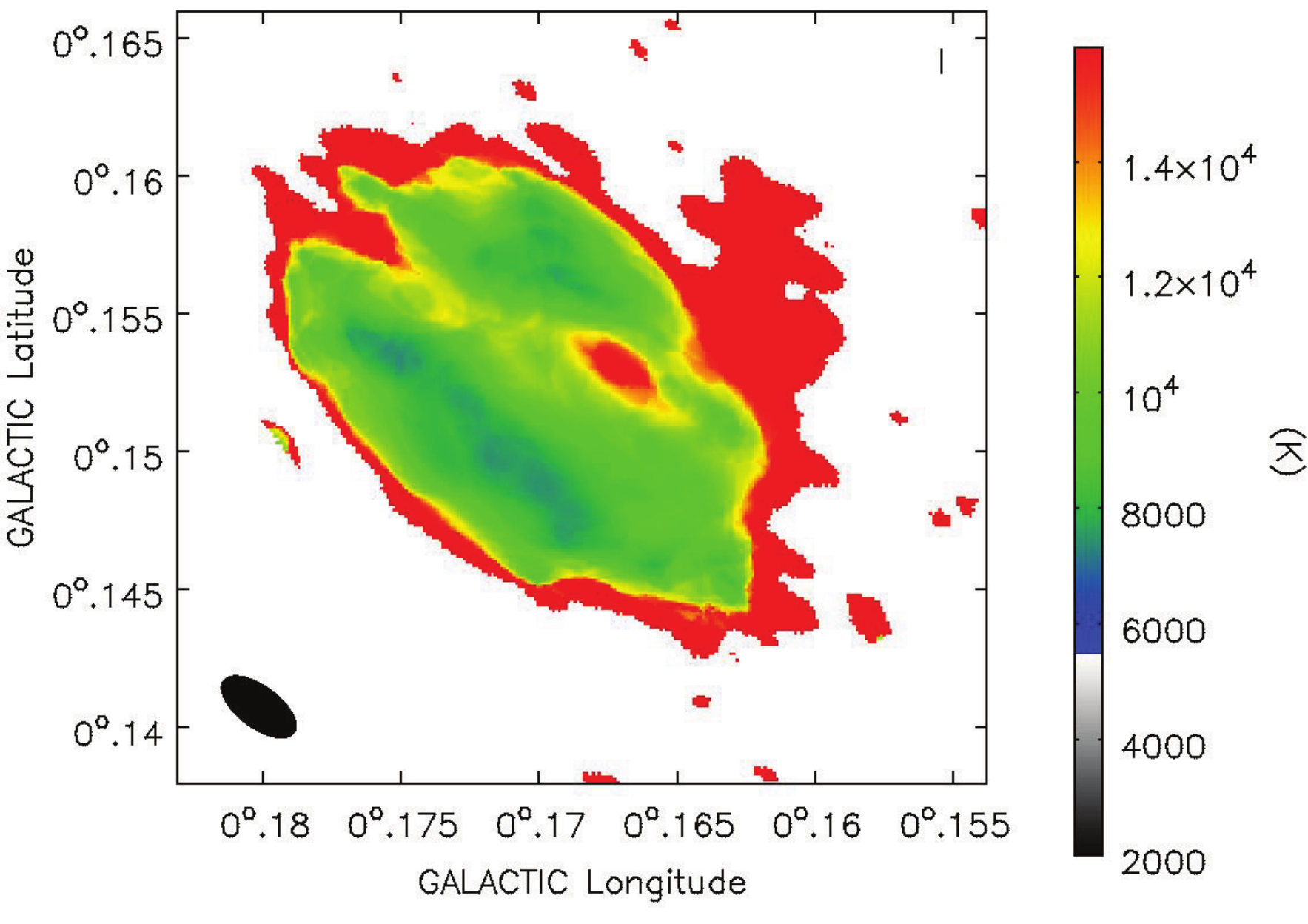} 
\includegraphics[scale=0.3, angle=0]{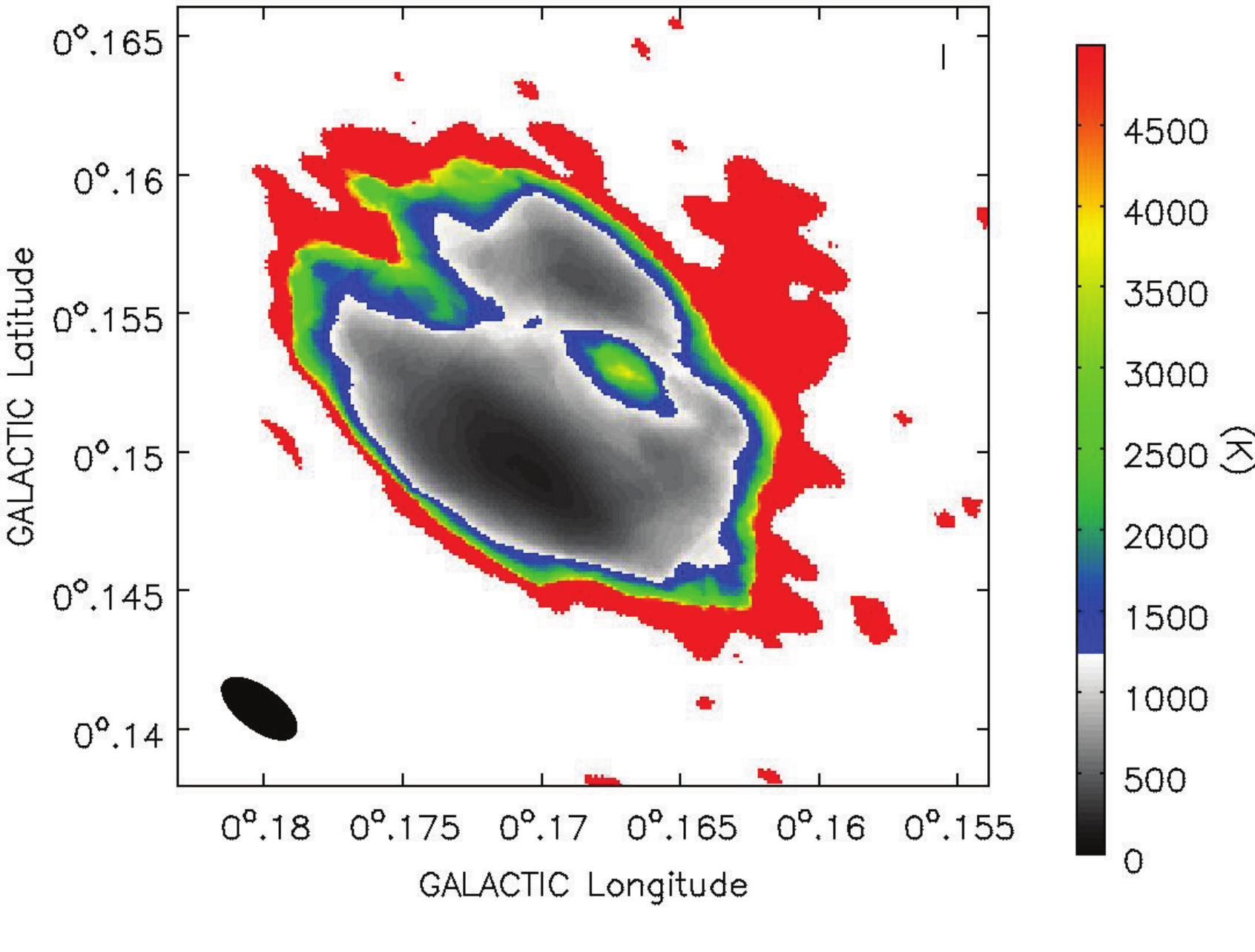}
\caption{
{\it (Top a)}
the ${\mathrm T_{e}}^*$ derived from RRL and continuum data observed with the VLA
in C and D array configuration at 10 GHz, assuming the electrons are under the LTE condition and 
the density ratio of helium to hydrogen is 0.07.
{\it (Bottom  b)}  $\sigma_{{\mathrm Te}^*}$, the errors
of the ${\mathrm T_{e}}^*$ image derived from 1$\sigma$ rms noises of the RRL and continuum images. 
The synthesized beam is marked as a filled ellipse in the bottom left of the plots. 
}
\end{figure}


To summarize, a number of kinematic arguments support the interaction picture. One is the evidence for the highest velocity ionized gas with a 
velocity gradient along the direction of the linear feature, as noted to   the Northwest of the ring. This feature is disjoint from the HII 
region 
likely to be ablated from the southern shell of G0.17+0.15 and 
accelerated in the direction  of the  filaments. Second, the evidence for stronger continuum and recombination line intensities, higher 
radial 
velocity and  higher velocity 
dispersions in  the southern shell of the ring when compared to the northern shell, all 
 suggest the possibility that the gas in the southern shell is 
kinematically disturbed as a 
result  of the interaction with nonthermal radio filaments.  
The above morphological and kinematic  arguments are plausible, however,  we have no 
direct proof of the 
interaction picture.




\subsubsection{Electron temperature}

Using Equation (1) of \cite{zbm2010}, from the  RRL (${\mathrm I_{RRL}}$) and 
continuum (${\mathrm I_{HII}}$) images, a distribution of the electron temperature (T$_{e}^*$) 
was computed (Fig. 8a),  assuming the ionized gas is under LTE conditions \citep[e.g.][]{wilson2009}.  
Based on the rms noises of RRL and continuum images, an error in ${\mathrm T_{e}^*}$ was also computed (Fig. 8b).
The 
typical values of 
${\mathrm T_{e}^*}$ 
are in the range between
7 and 10 $\times10^3$ K, with a 10 to 20\% uncertainty within the HII region.
The contamination from the synchrotron radiation (${\mathrm I_{syn}}$) leads 
an uncertainty in the
${\mathrm T_{e}^*}$ calculation.
Within the HII region with higher thermal electron density, 
the error due to the extended synchrotron emission is at a level of 10\%,  which is not included in
image of ${\mathrm T_{e}^*}$-error. 
In the diffuse area surrounding the HII region, for example around the shell
of the HII region and the southwest (G0.162+0.156), the ratio of ${\mathrm I_{syn}}$/I$_{\mathrm HII}$ drastically increases.
The derived high values of  
${\mathrm T_{e}}^*$ in these regions are upper limits because significant contribution from
synchrotron emission in the measured continuum flux density. Thus, an accurate measure of the electron temperature in the region where highest 
velocity is noted  is  not possible.       


\begin{figure}
\centering
\includegraphics[scale=0.8,angle=0]{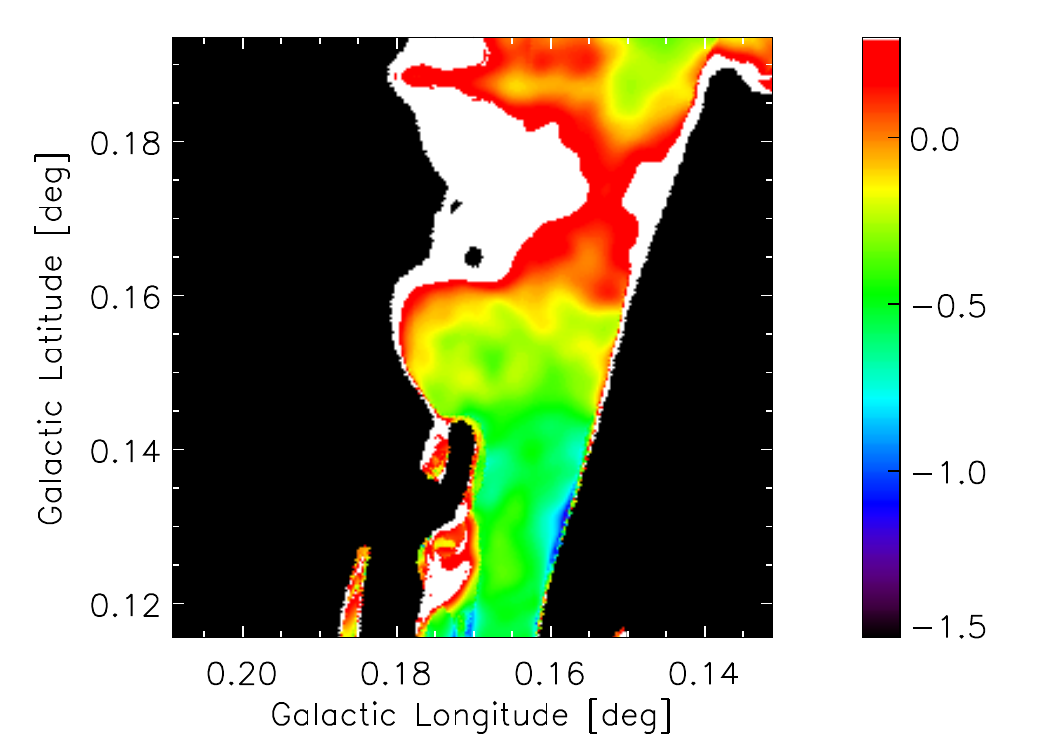}\\
\includegraphics[scale=0.8,angle=0]{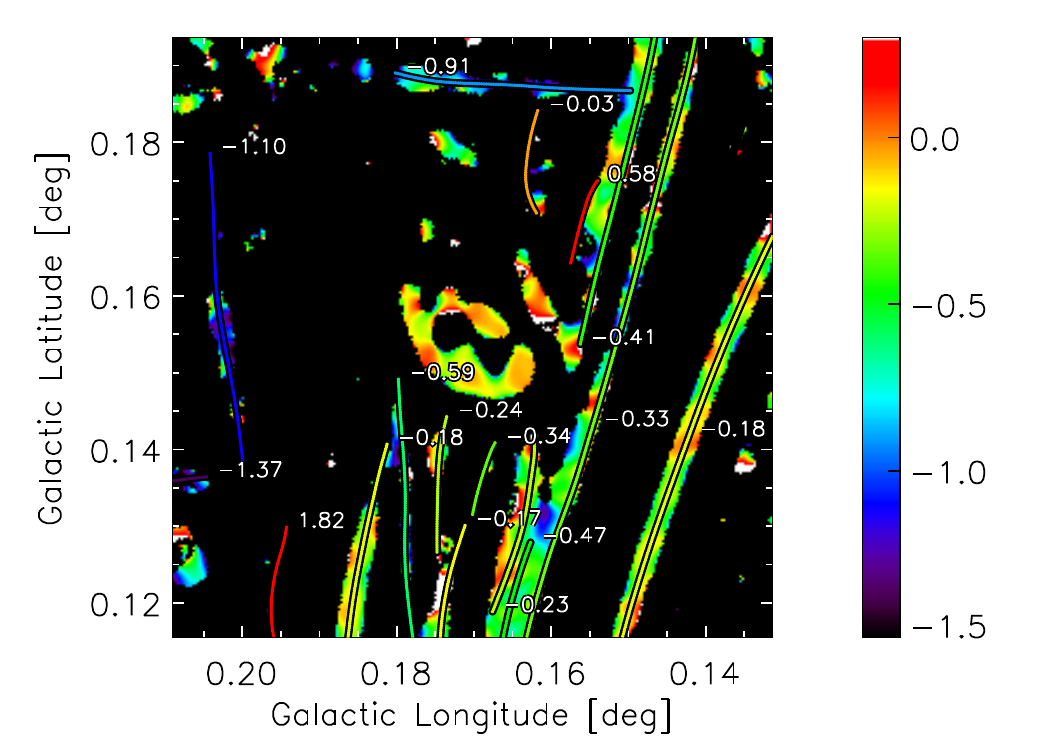}
\caption{
{\it (Top a)}
Spectral indices  of G0.17+0.15, nonthermal radio filaments  
using non-filtered  data cubes centered at 1.28 GHz.  White regions show $\alpha > 0.3$ whereas black regions
had  no sufficient data  and were  masked. 
{\it (Bottom b)} 
Similar to (a) except filtered data cubes were used. 
The spectral index image using  filtered data cubes,  
is overlaid with the median spectral index traced along each detected filament.
The scale bar to the right shows the spectral index values from -1.5 to +0.3. 
}
\end{figure}


\subsubsection{In-band spectral Index ($\alpha$) at 1.28 GHz}

Figure 9a,b show spectral index ($\alpha$) images, where the flux density is  $S_\nu\propto\, 
\nu^\alpha$, using unfiltered and filtered MeerKAT channel maps, respectively. The low-resolution ($8''$) 
spectral index image  in Figure 9a shows that the ring has a flat spectrum, $\alpha\sim-0.1$, 
consistent with being a thermal source \citep{heywood22}. Diffuse emission to the north of 
G0.17+0.15 also 
shows the spectrum of thermal  emission. 
The spectral indices  of the nonthermal filaments surrounding the HII 
region are  fairly flat as well. 
Figure 9b shows spectral indices derived from filtered images. With backgrounds subtracted, no spectral indices 
can be calculated for broad diffuse regions. The median spectral indices for the detected  nonthermal filaments are overplotted.
The filaments at the western edge of the ring have steeper 
spectral indices, $-0.5 < \alpha < -0.3$ than the isolated, short westernmost filament G0.146+0.14 with 
$\alpha\sim-0.18$.  Altogether, the filaments of the Radio Arc and its extension have flatter median 
spectral index values than those found in the Galactic center radio bubble, which have 
mean spectral index  $\alpha\sim-0.83$ \citep{zadeh22}. 

\begin{figure}
\centering
\includegraphics[scale=0.3, angle=0]{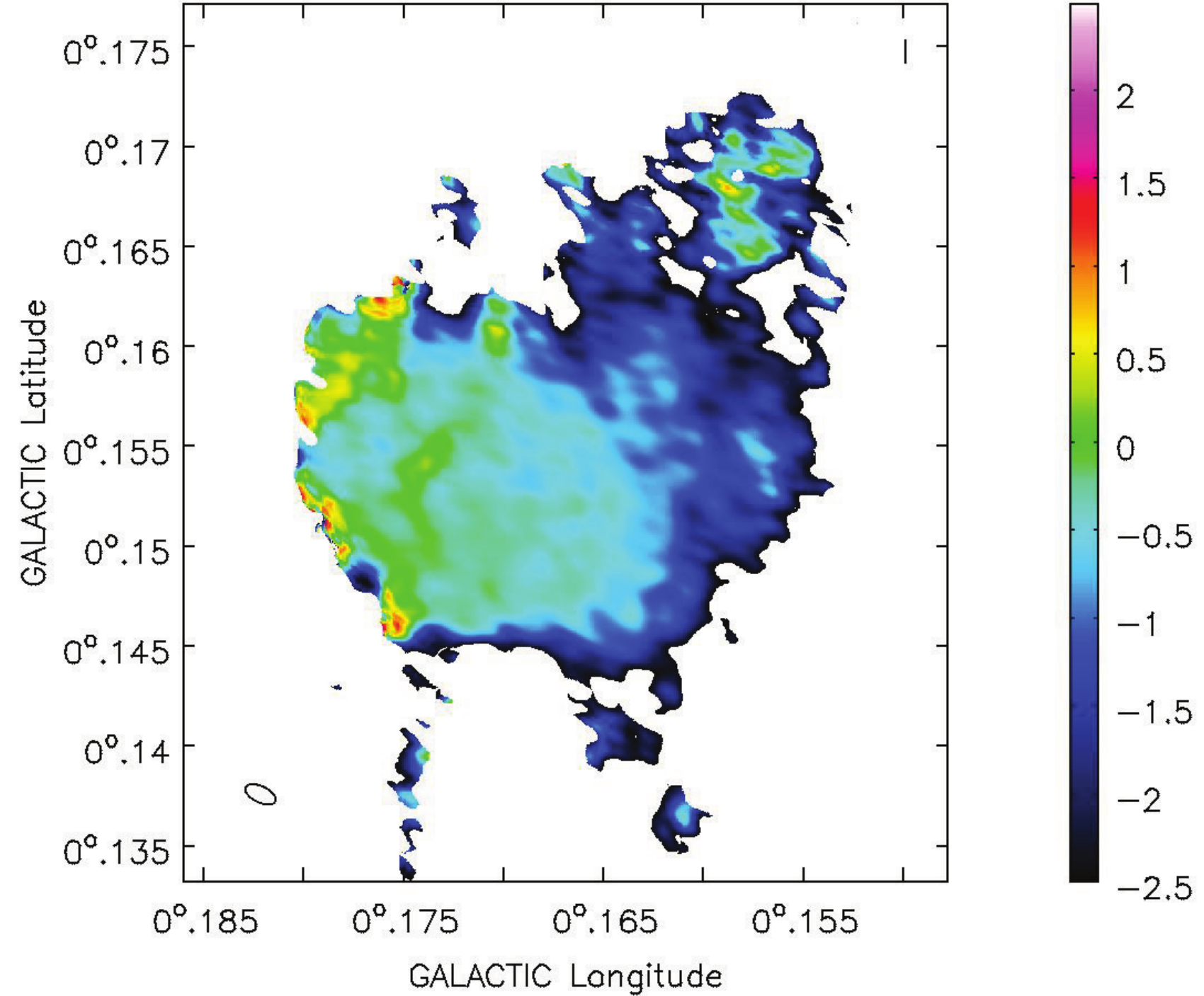}
\includegraphics[scale=0.3, angle=0]{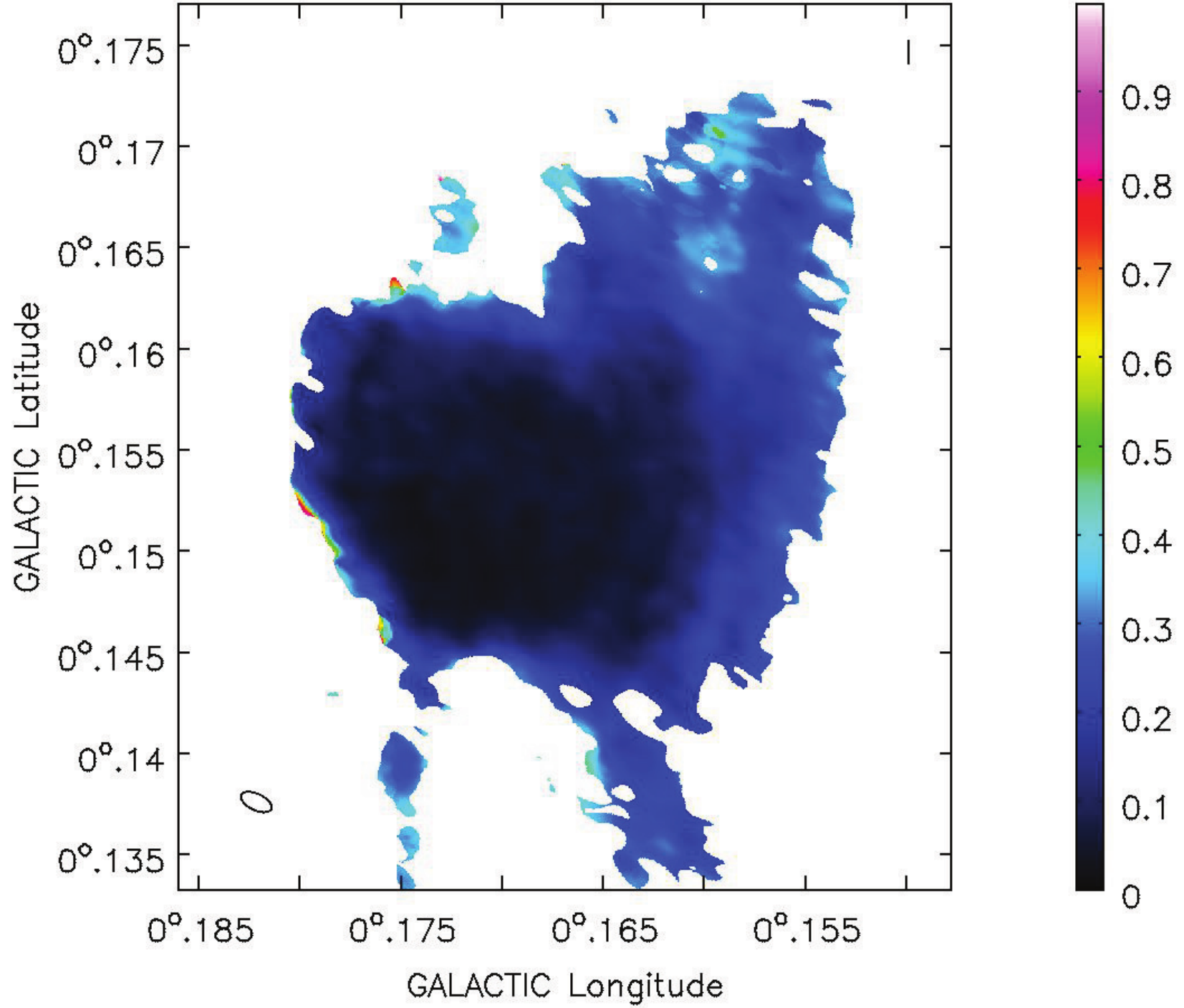}
\caption{
{\it (Left a)}
An image of in-band spectral index $\alpha$  computed from the VLA X-band data observed 
in the C- and D-array configurations. 
{\it (Right b)}
The corresponding one $\sigma$ error  $\sigma_\alpha$.
The beam size is FWHM = 5.9''$\times$2.9'' (6$^\circ$). A  5$\sigma_\alpha$ cutoff is applied for both images.
The display ranges are [-2.5, 2.5] for the spectral index $\alpha$ and [0, 1] for the 1 sigma uncertainty of the spectral index.
The synthesized beam is marked as an open ellipse in the bottom left of the plots.
}
\end{figure}



\subsubsection{In-band spectral index at 10 GHz}

Using the eleven sub-band images of X-band, we computed a set of 
$\alpha_{ij}$ 
images. An in-band average of $\alpha_{ij}$ is computed with the WASC algorithm. Figure 10  shows a result from WASC for
the distribution of spectral index. The relatively lower resolution image, good for sampling the diffuse emission, 
shows a significant gradient in $\alpha$ across the HII region. A 5$\sigma_\alpha$ filter has been 
applied to the $\alpha$
images. The steeper spectrum of the emission to the west of G0.17+0.15 could be the result of the 
contamination by diffuse nonthermal emission. 

\subsubsection{Rotation Measure (RM)}

Single dish observations of the entire 
eastern edge of the radio bubble with an extent of 2$^\circ$ have determined the rotation measure distribution,  
inferring that the intrinsic direction of the magnetic 
field traces vertical filaments perpendicular to the Galactic plane \citep{inoue84,seiradakis85,tsuboi86}. 
The rotation measure is an 
integrated  product of the electron density and the magnetic field component along the line of sight.  
 RM  values are  negative over the entire two degrees of the Radio Arc and its northern and southern extensions 
with the exception of the region surrounding G0.17+0.15,  where the RM is positive. The RM changes from a range 
of values between -500 and -1000 rad m$^{-2}$ to 500 and 1000 rad m$^{-2}$ at the western edge of G0.17+0.15 and 
its northern extension \citep{tsuboi86}.  This change coincides with the region where the spectral index values 
show both thermal and nonthermal emission. It is possible that entrainment of the G0.17+0.15 cloud injects 
thermal electrons in the positive latitudes of the Galactic center Lobe, changing the spectral index and 
stretching the magnetic field lines, thus making the sign of the RM change. 
The magnetic field direction could change  from being slightly  tilted away from the line of sight to being slightly tilted toward the line of sight. 
 For a typical 100 $\mu$G magnetic 
field strength, and a 2-pc path length, this would only require electron density $n_e > 10$ cm$^{-3}$. 
The  RM change  and the magnetic field bending near G0.17+0.15 
provide additional support for the interaction of nonthermal filaments with G0.17+0.15. 

\begin{figure}
\centering
\includegraphics[scale=0.8,angle=0]{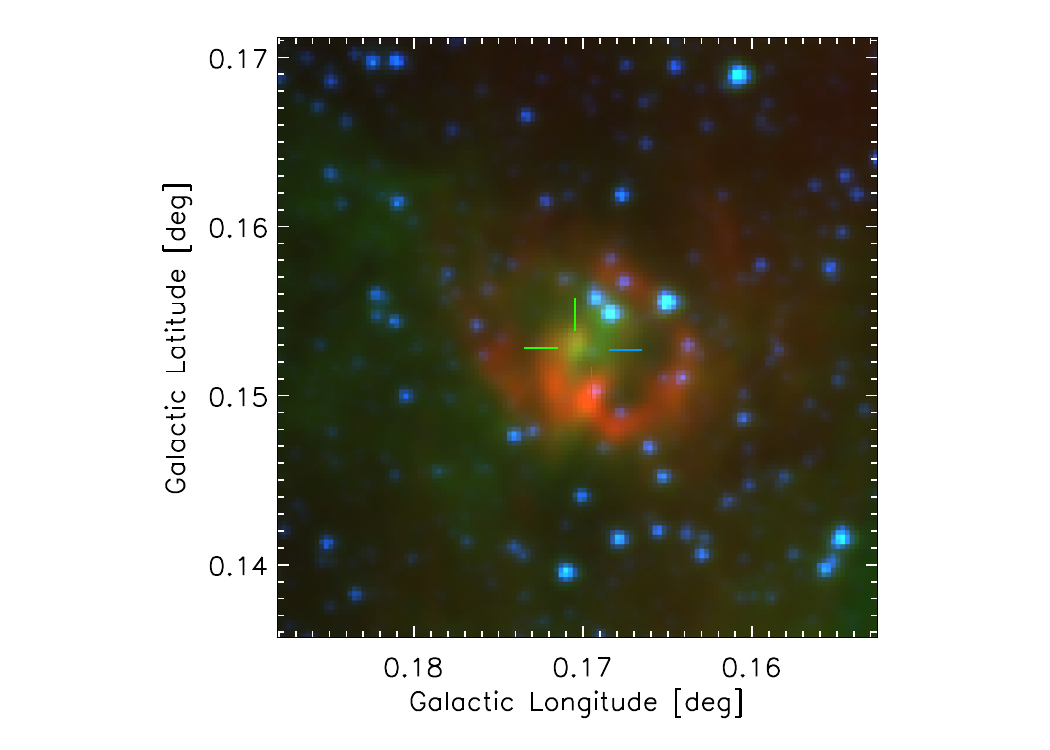}
\centering
\includegraphics[scale=1.6,angle=0]{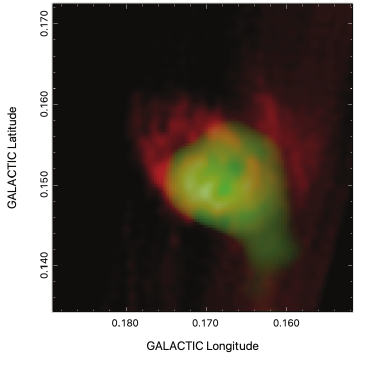}
\caption{
{\it (Top a)} 
A composite RGB color image of G0.17+0.15  is shown at 20cm (R), 8 $\mu$m (G) and 5.8 $\mu$m (B). 
The radio source  close to the center  of the HII region  has a diffuse 8$\mu$m counterpart. 
Green and blue tick marks indicate the locations of the two possible
central IR sources that are discussed in the text.
{\it (Bottom b)} 
MeerKAT radio images of total intensity (red, as in Figure 11a) at 4$''$ resolution, overlaid with the residual 
radio emission in green, at 8$''$ resolution, after removing all of the radio emission with $\alpha=0$, as described in the text.
}
\end{figure}

\begin{figure}
\centering
\includegraphics[scale=0.7,angle=0]{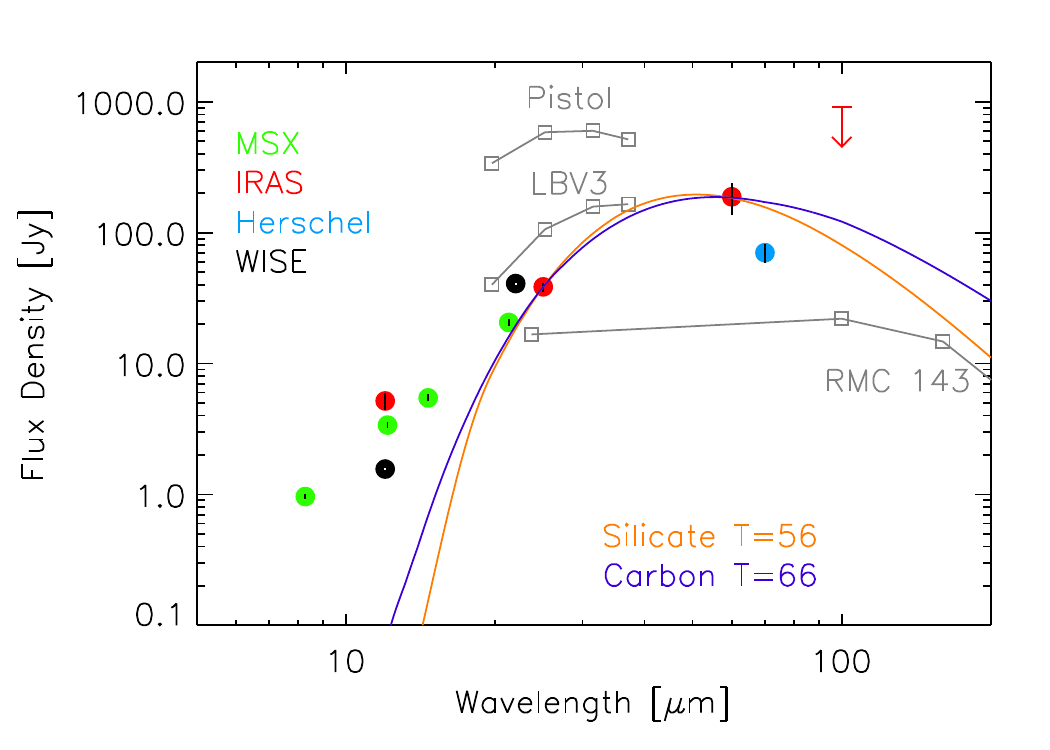}
\includegraphics[scale=0.7,angle=0]{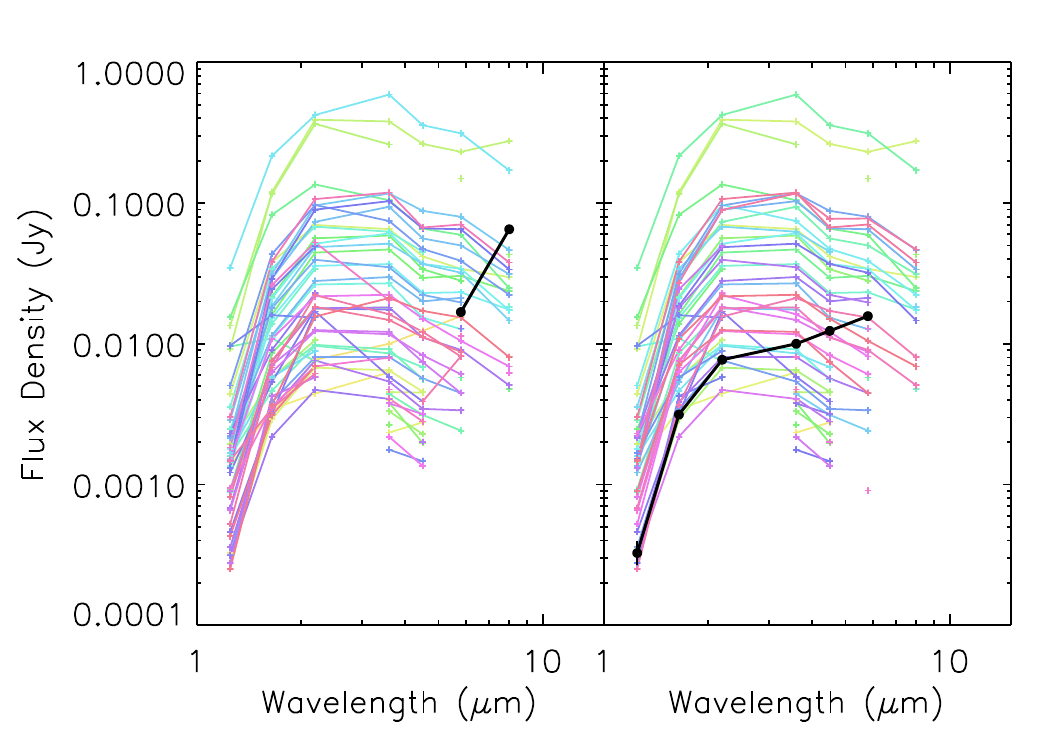}
\caption{
{\it (Top a)}
The SED of G0.17+0.15  as taken from various catalogs where the source is essentially unresolved 
(point-like sources). In the WISE catalog, the source is resolved into multiple components, but the one plotted is 
dominant at 12 and 22 $\mu$m. A single-temperature fit to the IRAS 25 and 60 $\mu$m data points is shown for 
either silicate or amorphous carbon dust grains. Gray SEDs for comparison are those of the shells around 3 LBV 
stars \citep{lau14,agliozzo19}.
Error bars are added to the G0.17 measurements. The error bars are generally smaller
than the plotted symbols (solid circles).
{\it (Bottom b)}
The heavy black SED in the left panel is that of the diffuse central source (indicated by green tick marks in Fig. 11a) as listed in 2MASS and 
Spitzer/IRAC catalogs. The other SEDs plotted are the field stars within a $2'$ radius (randomly colored for clarity), showing that no other nearby 
sources have a similar SED. The black line in the right panel shows that the SED of the bluer central point-like source (indicated by blue ticks in 
Fig. 11a) has a strong excess at $>3.6$ $\mu$m that is not present in any of the field stars.
}
\end{figure}

\subsubsection{IR and spectral tomography images}

Figure 11a shows an RGB color image of G0.17+0.15 at 20cm, 8$\mu$m and 5.6$\mu$m.  The Spitzer IRAC images show 
diffuse IR emission extending over the entire radio structure, with a rising spectrum between 5.8 and 8 $\mu$m, 
so red in IR color.

To see whether there was a radio counterpart to the IR structure, we applied  the spectral tomography technique 
of \cite{rudnick97} to the MeerKAT data.  In this technique, a series of residual images is created by removing 
all of the emission with a specific spectral index ($\alpha_t$).\footnote{To construct the residual images, we 
subtracted an image of the average of $8''$ MeerKAT channels 11, 12 and 14 (effective frequency 1542.4~MHz) from 
the average of channels 0-3 (962.85~MHz), with scale factors corresponding to a range of spectral 
indices, -1.5 $<\, \alpha_t < 0.35$.} 
In the residual images for $\alpha_t~=~0\pm0.05$, the ring is no longer visible, 
so the spectrum of the ring is essentially flat.  The residual emission is shown in Figure 11b.  
 It peaks up 
inside of the ring, and is therefore somewhat steeper, and  has a location and extent similar to the diffuse IR feature in Figure 11a. This diffuse 
radio 
component has a  spectral index of $\sim$-0.2, as determined from when it disappears from the tomography images.

The spectral indices of the ring and its interior are different, 
 suggesting different origin.  The interior 
emission could be produced either by shocked stellar winds from a stellar source embedded within the ring or by 
large-scale cosmic-ray-driven nuclear wind interacting with an HII region and diffusing into the HII region.

\subsubsection{SEDs of nearby sources}

The diffuse IR source that comprises the entirety of G0.17+0.15  is saturated in the Spitzer MIPS 24 $\mu$m image,  but has unsaturated mid-IR 
measurements 
from IRAS, MSX, WISE and Herschel (see Fig.  11a). To determine the energetics of this source, the 
IRAS 25 and 60 $\mu$m data points of the SED, were fit as either silicate or amorphous carbon dust grains at a 
single temperature. These fits, shown in Figure 11a, provide estimated dust temperatures of T=56 or 66 K, for 
silicate or carbon respectively. The implied dust masses are 0.10 \msol\, for silicate or 0.067 \msol\, for 
amorphous carbon. Assuming a dust/gas mass ratio of 0.0062, the implied gas masses are 16 and 11 \msol. Dust and 
gas masses may be higher if a cooler component is present, but the lack of detection at 100 and 160 $\mu$m 
suggests that there is not a significant colder component. The mid-IR luminosity of this source is $\sim 2\times 10^4$ \lsol.

Using higher resolution 2MASS and Spitzer IRAC data, we explored the SEDs of IR sources in the 2$'$ vicinity of the HII region. 
In addition to ubiquitous stars, mid-IR images also reveal emission associated
with G0.17+0.15, as seen in the composite radio and IR image in Figure 11a. 
There are two unusual IR sources in the center of G0.17+0.15. 
The first is slightly more extended than a point source and appears coincident with a compact radio peak. This source has very red IR colors and is 
only cataloged at 5.8 and 8 µm. It is the yellow-green blob,  pointed by a green tick mark on Figure 11a. 
Its SED is shown by the heavy black circles and line in  Figure 12a (left panel). 
The second unusual source appears point-like, and is displaced about 2$''$  to the West. It is a relatively faint bluish source, partially merged 
with the 
preceding source, in Figure 11a. It is only detected at 1.25-5.8 $\mu$m. 
Coincidentally, its 5.8 µm flux density is very similar to that of the 
preceding 
source. Its SED is shown by  circles and black  line lines in Figure 12b (right panel)  that shows an unusual excess in emission at 3.6-5.8 µm 
compared to all of 
the other stellar 
sources in the 2' vicinity. 
These are the (a) and (b) choices discussed in  the next paragraph.


The presence of a distinctive central source (perhaps two) suggests that that the ring of 
G0.17+0.15 might not be 
an HII region associated with star formation, but rather the result of 
of mass-loss from an evolving high mass 
star. Shells with a somewhat similar appearance and SEDs are seen around the LBV stars such as the Pistol star and 
LBV 3 in the Galactic center and RMC 143 in the LMC \citep{lau14,agliozzo19}. There are two choices for the 
responsible source: (a) It may be coincident with the radio source having  a  diffuse,  red,  IR counterpart. The 
IR would 
suggest the source is deeply embedded in a  cloud, making it perhaps unlikely that the star's UV radiation 
could escape to ionize the ring. In this scenario, the Arches cluster may ionize the HII region externally. 
This is analogous to the Pistol nebula being photoionized externally by the Quintuplet cluster star (\cite{lau14}, and references therein). The 
implied mass of 11-16 \msol\, lost by the star is 
somewhat 
high, but the estimate could be reduced if the dust-to-gas mass ratio in the circumstellar material is higher 
than typical for the ISM. (b) The near-IR point source may be  the source of ionization. There is far less 
circumstellar material here. UV radiation could escape to ionize the ring  and may also provide an external 
heating source for dust in the nearby diffuse IR source.

In either case, an additional constraint can be placed by estimating the number of Lyman continuum photons needed 
for a stellar source to ionize the gas. Assuming an electron temperature of 8000K, and a distance of $\sim8$ kpc from us 
and a flux density of $\sim0.3$ Jy, the total estimated Lyman continuum rate is $1.6\times 10^{48}$ photons 
s$^{-1}$. 
Assuming that the emission is produced by a single ZAMS star, a
spectral type of O8-9 would produce sufficient ionizing photons to drive the  
emission \citep{panagia73}.  
If the star is an  evolved supergiant, then a later
spectral type of B0 can produce the same ionizing flux. The total stellar
luminosity is $\sim 5\times 10^4$ \lsol, for a ZAMS star (or up to $\sim 3\times 10^5$ 
for a supergiant), sufficient to produce the dust luminosity
estimated above. 
On the other hand,  it is possible that G0.17+0.15 is formed in the clumpy large-scale molecular cloud. 
If there is such a physical association, the high-velocity of ionized gas traces orbital motion of gas clouds in the Galactic center.

\section{Discussion}


An implication of the physical association of high-velocity,  diffuse HII gas and molecular clouds with the 
eastern and western Lobes, respectively, is that 
the Lobes, the radio bubble, and NRFs  are located at the Galactic center. 
Two   features that need to be addressed are  whether the  HII region G0.17+0.15 is interacting with 
NRFs,  as well as being  accelerated 
by the ram pressure of cosmic-ray driven outflow directed along 
the nonthermal filaments. 
The radio bubble hosts hundreds of mysterious magnetized radio filaments, the largest 
concentration of which is  the radio Arc. The decline in the number of filaments as a 
function of Galactic longitude and latitude are consistent with the size of the radio bubble, suggesting 
a causal association with the origin of the filaments \citep{heywood19}.  One scenario for 
the origin of the bipolar radio/X-ray features involves a  high cosmic-ray pressure driving large-scale winds and 
expanding the medium away from the Galactic plane \citep{zadeh19}. The cosmic ray ionization rate in the 
Galactic center is two to three orders of magnitudes higher than in the Galactic disk, implying that  the 
cosmic-ray pressure is sufficient to launch a nuclear wind from the Galactic center \citep{zadeh19}.

The above interpretation is not unique. It is possible that 
 infall motion of the cloud and/or HII region from
its high-latitude location towards the Galactic center, can explain some of the observed features. 
In this picture, G0.17+0.15 is   formed in the clumpy large-scale 
molecular cloud.
However, if G0.17+0.15 were part of a larger cloud that is falling into the disk of the Milky Way, then it is an unfortunate coincidence that it 
is superimposed on the Radio Arc at relatively low latitudes of $\sim0.15^\circ$, and that there are no other similar structures elsewhere in the 
CMZ.


 Here we only focus on plausibility of the the nuclear wind.
We assume that the interacting  nuclear wind can explain 
 the cause of morphological and kinematical asymmetry in G0.17+0.15. 


\subsection{Asymmetric ablation of the western edge of the ring}
 
We made a number of morphological and kinematic arguments of an asymmetry in the radial velocity field and brightness distribution across the width of
G0.17+0.15. The asymmetry suggest that the ionized gas could be ablated by an external ram pressure of the nuclear wind as traced by a larger
concentration of the filaments at the western edge of the HII region.  The electron density $n_e$ from the western side of the ring is estimated to be
$\sim300\, \rm cm^{-3}$ using an intensity of $\sim7$ mJy per 4$''$ beam at 1.28 GHz.  The mass of ionized gas in G0.17+0.15 is then $\sim20$ \msol, 
consistent with the dust mass inferred above.
 
Using a velocity difference of $\delta v \sim 15$ \kms\ across the diameter of the ring, $L\sim1.5$ pc, and the ionized gas density of
$n_e\sim3\times10^2$ cm$^{-3}$, the needed ram pressure, is 500\,eV\,cm$^{-3}$, which is comparable to the $\sim10^3\, {\mathrm eV cm^{-3}}$ 
cosmic-ray driven wind
pressure needed to explain X-ray filled radio bubble \citep{zadeh19} and consistent with a scenario in which this wind is responsible for the 
ablation.   The thermal pressure of the ionized gas for $n_e = 300$ cm$^{-3}$ and 8000K is 450\,eV\,cm$^{-3}$, so the HII
region is likely to be confined by its surroundings.
 
In the above picture, the gas streamers on the western side of the HII region are stripped and assumed to have been uniformly accelerated from 0 to
$\delta v\sim\,35$ \kms\, over a length of $l\sim1.4$ pc.  The acceleration is $0.5\times\delta v^2 / l$ over a time scale $2\times l / \delta v$,
which is roughly $8\times10^4$ years. The estimated column density of the ablated material $\sim5\times10^{20}$ cm$^{-2}$, implies that the required
ablating ram pressure is $\sim700$\,eV\,cm$^{-3}$, which is also comparable to the $\sim10^3$\,eV\,cm$^{-3}$ cosmic-ray driven wind pressure.




\section{Summary}

We studied G0.17+0.15, lying in the eastern (positive Galactic longitudes and high Galactic latitudes)  Lobe of the Galactic center, respectively. 
Although there are other ideas that can explain its  structure, the kinematics  and origin,  
our focus was mainly on the HII region G0.17+0.15,
because of its high radial velocity $\sim130$ \kms. Based on its morphology, kinematics and polarization 
measurements, we provided  arguments for the interaction of G0.17+0.15 with a number of nonthermal radio filaments which themselves are 
considered to be accelerated by a cosmic-ray driven wind inflating the Galactic center radio bubble. As such, the properties of G0.17+0.15 are 
consistent with entrainment. Velocity gradients along the linear features to the west of G0.17+0.15, as being ablated by the cosmic-ray driven nuclear 
wind. We also found two IR sources within G0.17+0.15. The presence of a distinctive central source may be an indication of 
mass loss from an evolving high mass star. The SED of this red source appears to be similar to that of LBV stars 
such as the Pistol star and LBV 3 in the Galactic center.  

The above ideas  are not unique in explaining G0.17+0.15. It is possible that G0.17+0.15 is a site of star formation in high-velocity 
orbiting molecular 
clouds and that massive stars are formed in the clumpy molecular environment of the Galactic center.  Future high-resolution radio and infrared 
observations will be  able to test the physical association of G0.17+0.15 and large-scale molecular clouds and identify whether massive or evolved 
stars power the HII region.





\section{Data Availability}

All the data including   MeerKAT  that we used here are available online and are not proprietary.
We have reduced and calibrated these data and these are available if  requested.

\section*{Acknowledgments}
This work is partially supported by the grant AST-2305857 from the
National Science Foundation. Work by R.G.A. was supported by NASA under award number 80GSFC21M0002.
The National Radio Astronomy Observatory is a facility of the National Science Foundation
operated under cooperative agreement by Associated Universities, Inc.









\input{table2_final.tex}

\section{Appendix} 
\subsection{Stacking multiple $\alpha-$transition lines from an hydrogen-like atom}

The image cubes of eleven RRLs were stacked in order to enhance the signal-to-noise ratio.
We investigated a few algorithms in stacking the RRL images for optimizing the sensitivity. 
Here is a description of how to weight the individual RRL image cubes in the stacking process
concerned in this paper:
$$
\overline{\rm I_{\rm RRL}} = {\bf wt}\cdot {\bf I_{RRL}}, 
$$
\noindent where ${\bf  I_{RRL}}$ is a 1-D image matrix and its element is a RRL image cube at an $\alpha$-transition
corresponding to a principle quantum number ({\rm n}) of an hydrogen-like atom:
$$  {\bf I_{\bf RRL}} = 
\begin{pmatrix}
{\rm  I_{RRL}}[0] \\ {\rm I_{RRL}}[1]\\ \vdots\\ {\rm I_{RRL}}[{\rm m}]
\end{pmatrix}, $$
\noindent and the weight matrix:
$$  {\bf wt} = 
\begin{pmatrix}
{\rm  wt}[0], {\rm wt}[1], \dots, {\rm wt}[{\rm m}]
\end{pmatrix}. $$
In the case of
the X-band VLA data, ${\rm m}=10$ and $i$ is a serial number of the {\bf I$_{\rm RRL}$} elements, 
where $i$ = 0, 1, ..., and 10, 
corresponds to $n$ = 92, 91, ... and 82;
and wt[$i$]  is a statistical weight used in the process 
of stacking the multiple RRLs. The final resultant RRL
image cube $\overline{\rm I_{\rm RRL}}$ is a dot-product of the weighting matrix ${\bf wt}$ and
the RRL image  matrix ${\bf I_{RRL}}$. Three different weighting methods were considered in this paper:\\
(1) equal weighting or $${\rm wt} [i] =  \left(\sum i\right)^{-1};$$ 
(2) weighted by the reciprocal value ($\sigma_i^{-2}$) of a variance determined from a RRL image cube,
$$ {\rm wt} [i] = {1\over\sigma^2_{i}} \left(\sum {1\over\sigma^2_{i}}\right)^{-1}; $$
and (3) weighted by a quadratic power of an S/N ratio, 
$$ {\rm wt} [i] = \left({\rm I}^p_{i}\over\sigma_{i}\right)^2 
\left(\sum {\left({\rm I}^p_{i}\over\sigma_{i}\right)^2}\right)^{-1}. $$
In above equations, $\sigma_{i}$ and ${\rm I}^p_{i}$ denote a rms noise and peak line 
intensity of the $i$th element in the RRL image cube matrix. We found that weighting with 
a quadratic power of an S/N ratio or $\left({\rm I}^p_{i}\over\sigma_{i}\right)^2$  gives the best sensitivity 
among the three stacked image cubes.  The final RRL image cube computed from method (3) is improved
in sensitivity by $\sim$ 10\%  and $\sim$ 5\% as compared to the one produced with 
an equal weighting function and a weighting with the reciprocal value of a variance, respectively. 
The sensitivity of the stacked RRL image cube is improved by a factor of 4 as compared to that of
 an individual RRL cube.
Thus, we adopted the stacked RRL image cube produced by weighting with  a quadratic power of an S/N ratio. 

\subsection{VLA In-band Spectral Index}

Both the VLA's  X-band and C-band data have  a broad frequency coverage of  4 GHz in bandwidth ($BW$),
providing a large ratio of $\left[BW\over\nu_0\right]$ = 0.67 and 0.4 at 6 and 10 GHz, respectively.
Wideband imaging with the Taylor expansion technique \citep[e.g.][]{raucornwell2011} has 
constructed a multi-frequency synthesis (MFS) image of the continuum emission. 
This has been achieved 
 from a radio source via  
Taylor expansion's first term  or TT0, providing deep images with a high dynamic range.
The second term Taylor expansion (TT1) deals with the slope of  spectrum distributed across a wideband, 
from which one can derive images for distributions of spectral index. 
However, this technique is challenging because  (1) cleaning sidelobes for TT1 requires high signal-to-noise ratio 
due to the TT1 beam being  orthogonal to that of TT0 \citep{ccw1990} and (2)
the wideband coverage makes the linear expansion inappropriate for a source with a non-linear spectrum. 
A  higher order expansion is needed to interpret the spectrum curvature. 
Thus, the  reliability for extended
sources was uncertain, often owing to an inadequate signal-to-noise ratio. 
On the other hand, the classical method uses two images I$_1$ and I$_2$ produced from data observed  
at $\nu_1$ and $\nu_2$ in a narrow band ($\Delta\nu << \nu_{1}$  and 
$\Delta\nu << \nu_2$)
following a formula: 
$$\alpha = {\rm Log\left(\it I_1/ I_2\right) \over Log\left(\rm \nu_1/ \nu_2\right)}.$$ 
\noindent The VLA was designed via array configurations and receiver bands to provide
options for users to observe a source in multiple observing 
runs  resulting in an equivalent uv-sampling
at two separate band frequencies. In principle, the narrow-band approach  reduces the uncertainties
in determination of spectral index for strong sources. 
Given a wideband observation, numerous sub-bands produced in a single observing run,
from which one can produce a large number of spectral index $\alpha$ 
images from a combination of a frequency pair ($\nu_{i}$, $\nu_{j}$), 
where $i$ or $j$ is a serial number of sub-bands, an integer from 
0, 1, 2, ... to $n$, with a total number of $n+1$ sub-bands. A total number of sub-bands is $n+1$.
For the VLA data taken at X- and C-bands, 
the corresponding $n+1$  is 12 and 32. An assemble
of spectral index $\alpha_{ij}$ images can be computed using above formula as well as
their errors  $\sigma^\alpha_{ij}$ based one the rms noises of
images and a separation of a frequency pair ($\nu_{i}$, $\nu_{j}$). 
We note that both $\alpha_{ij}$ and  $\sigma^\alpha_{ij}$
are two dimensional images with RA and Dec in the $x$ and $y$ axes. For example, 
for the complex HII region G0.17+0.15 at X-band, $\alpha_{ij}({\mathrm x}, {\rm y})$ was derived from a pair of 
5,000  spectral images I$_{i}({\mathrm x}, {\mathrm y})$
and I$_{j}({\mathrm x}, {\mathrm y})$. Instead of fitting to the higher order terms in Taylor expansion,
the spectral index $\alpha$ using the classical method involves only the first term 
(TT0) of the Taylor expansion 
products. In addition, the bandwidth of a sub-bands $\Delta\nu_{i}$ appears to be  small enough,
so a linear approaching is valid for the slope of a radiation spectrum.  For example, $\Delta\nu_{i}$
is 64 and 128 MHz ($\Delta\nu/\nu =$ 0.006 and 0.02) for the VLA X-band and C-band data.
We also note that the sub-band data taken in a wideband observation are subject to 
the frequency-dependent issues
in aperture synthesis interferometer imaging, such as primary beam attenuation, short-baseline cutoff and frequency smearing
as well as radio frequency interference.
Those effects are accounted for  in the process of constructing  $\alpha$ as well as by 
computing a weighted-average of  a sub-band combination (WASC) of  frequency-pair images $\alpha_{\rm ij}$. The statistical errors  reflect the uncertainty of the algorithm.


\end{document}

%% file: table2_final.tex
\begin{deluxetable}{llllllll}
\tablecolumns{8}
\tablewidth{0pt}
\tablecaption{\leftskip=0.1em\label{tab:table1} Gaussian fits to integrated RRL  spectrum per 1.6$'$ beam}
\tablehead{
\colhead{} &
\multicolumn{2}{c}{{Center Position}} & & 
\multicolumn{3}{c}{Gaussian Fits}  &
\\
\cline{2-3}
\cline{5-7}

\colhead{Name} &
\colhead{$l$ } &
\colhead{$b$ } & &
\colhead{$v\pm1\sigma$} &
\colhead{peak flux$\pm1\sigma$} & 
\colhead{FWHM$\pm1\sigma$}&
\colhead{comments}\\

\colhead{} &  
\colhead{(deg)} &
\colhead{(deg)} & &
\colhead{(km/s)}&
\colhead{(mJy)}&
\colhead{(km/s)}&
\colhead{}
}
\startdata
  Center &      0.1700 & 0.1502 & & 3.1$\pm0.03$  & 4.2$\pm0.05$    & 19.9$\pm0.16$ & Hen$\alpha (87< n < 93)$\\   
   -     &           - &     -  & & 128$\pm0.37$   & 16.4$\pm0.78$    & 25.0$\pm2.05$ &Hn$\alpha (87< n < 93)$\\ 
  South  &      0.1700 & 0.1127 & & -12.8$\pm0.12$ & 18.8$\pm0.11$   & 40.1$\pm0.67$ &Hn$\alpha (87< n < 93)$\\   
  North  &      0.1700 & 0.1877 & & 7.5$\pm0.42$  &  8.0$\pm0.23$   & 12.6$\pm1.26$ &Hn$\alpha (87< n < 93)$\\   
  East   &      0.2075 & 0.1502 & & 127.0$\pm6.18$   & 2.1$\pm0.075$    & 38.0$\pm34.3$ &Hn$\alpha (87< n < 93)$\\   
  West   &      0.1325 & 0.1502 & & x     & x       & x &background noise\\   
  Northwest &   0.1614 & 0.1564 & & 3.5$\pm0.41$  & 4.5$\pm0.13$    & 19.1$\pm0.3$ &Hen$\alpha (87< n < 93)$\\      
  -         &    -     & -      & & 131.0$\pm0.04$   & 14.1$\pm0.06$    & 24.3$\pm0.24$&Hn$\alpha (87< n < 93)$\
\enddata
\end{deluxetable}